\documentclass[aps, prd, twocolumn, amsmath, floats, floatfix, superscriptaddress, nofootinbib
]{revtex4-2}
\usepackage{siunitx}
\usepackage{epsfig}
\usepackage{amsmath,amsfonts,amssymb}
\usepackage[caption=false]{subfig}
\usepackage{verbatim}
\usepackage{float}
\usepackage{morefloats}
\usepackage[dvipsnames]{xcolor}
\usepackage{booktabs}
\usepackage[normalem]{ulem}
\usepackage{multirow}
\usepackage{makecell}
\usepackage{tabulary}
\usepackage{cancel}
\usepackage[T1]{fontenc}

\usepackage{hyperref}       
\hypersetup{
	colorlinks   = true, 
	urlcolor     = black, 
	linkcolor    = black, 
	citecolor    = black,  
}

\usepackage[capitalise]{cleveref}
\usepackage{comment}
\crefname{section}{Sec.}{Secs.}
\crefname{table}{Tab.}{Tabs.}
\crefname{figure}{Fig.}{Figs.}
\crefname{equation}{Eq.}{Eqs.}
\crefname{appendix}{Appendix\ }{Appendix\ }

\providecommand{\openone}{\leavevmode\hbox{\small1\kern-3.8pt\normalsize1}}

\DeclareSIUnit\parsec{pc}
\DeclareSIUnit\megaparsec{Mpc}
\DeclareSIUnit\solar{$\mathrm{M}_\odot$}

\definecolor{bostonuniversityred}{rgb}{0.8, 0.0, 0.0}

\usepackage{booktabs}
\usepackage[Q=yes,pverb-linebreak=no]{examplep}
\begin{document}

\title{\boldmath Deep-learning classification and parameter inference of rotational core-collapse supernovae\unboldmath}

\author{Solange~Nunes}
\thanks{Corresponding author}
\email{solangesilnunes@gmail.com}
\affiliation{Centro de F\'{\i}sica das Universidades do Minho e do Porto (CF-UM-UP), Universidade do Minho, 4710-057 Braga, Portugal}

\author{Gabriel~Escrig}
\email{gescrig@ucm.es}
\affiliation{Departamento de F\'{ı}sica Te\'orica, Universidad Complutense de Madrid,\\
Plaza de las Ciencias 1, 28040 Madrid, Spain}
  
\author{Osvaldo~G.~Freitas}
\email{ogf1996@gmail.com}
\affiliation{Centro de F\'{\i}sica das Universidades do Minho e do Porto (CF-UM-UP), Universidade do Minho, 4710-057 Braga, Portugal}
\affiliation{Departamento de
  Astronom\'{\i}a y Astrof\'{\i}sica, Universitat de Val\`encia,
  Dr. Moliner 50, 46100, Burjassot (Val\`encia), Spain}

\author{Jos\'e~A.~Font}
\email{j.antonio.font@uv.es}
\affiliation{Departamento de
  Astronom\'{\i}a y Astrof\'{\i}sica, Universitat de Val\`encia,
  Dr. Moliner 50, 46100, Burjassot (Val\`encia), Spain}
\affiliation{Observatori Astron\`omic, Universitat de Val\`encia,  Catedr\'atico 
  Jos\'e Beltr\'an 2, 46980, Paterna (Val\`encia), Spain}

\author{Tiago~Fernandes}
\email{tiagosfernandes@ua.pt}
\affiliation{Centro de F\'{\i}sica das Universidades do Minho e do Porto (CF-UM-UP), Universidade do Minho, 4710-057 Braga, Portugal}

\author{Antonio~Onofre}
\email{antonio.onofre@cern.ch}
\affiliation{Centro de F\'{\i}sica das Universidades do Minho e do Porto (CF-UM-UP), Universidade do Minho, 4710-057 Braga, Portugal}

\author{Alejandro~\surname{Torres-Forn\'e}}
\email{alejandro.torres@uv.es}
\affiliation{Departamento de
  Astronom\'{\i}a y Astrof\'{\i}sica, Universitat de Val\`encia,
  Dr. Moliner 50, 46100, Burjassot (Val\`encia), Spain}

\begin{abstract}


We test deep-learning (DL) techniques for the analysis of rotational core-collapse supernovae (CCSN) gravitational-wave (GW) signals by performing classification and parameter inference of the maximum (peak) frequency and the GW strain amplitude ($\Delta h$) multiplied by the luminosity distance ($D$) attained at core bounce, respectively, $(f_\textrm{peak})$ and $(D \cdot \Delta h)$. Our datasets are built from a catalog of numerically generated CCSN waveforms assembled by Richers~\textit{et al.}~2017. Those waveforms are injected into noise from the Advanced Laser Interferometer Gravitational Wave Observatory and Advanced Virgo detectors corresponding to the O2 and O3a observing runs. For a network signal-to-noise ratio (SNR) above 5, our classification network using time series detects Galactic CCSN GW signals buried in detector noise with a false positive rate of 0.10\% and a 98\% accuracy, being able to detect all signals with SNR~>~10. The inference of $f_\textrm{peak}$ is more accurate than for $D \cdot \Delta h $, particularly for our datasets with the shortest time window (0.25 s) and for a minimum SNR~=~15. From the calibration plots of predicted versus true values of the two parameters, the standard deviation ($\sigma$) and the slope deviation with respect to the ideal value are computed. We find $\sigma_{D \cdot \Delta h} = 52.6$~cm and $\sigma_{f_\textrm{peak}} = 18.3$~Hz, with respective slope deviations of 11.6\% and 8.3\%. Our best model is also tested on waveforms from a recent CCSN catalog built by Mitra~\textit{et al.} 2023, different from the one used for the training. For these new waveforms, the true values of the two parameters are mostly within the $1\sigma$ band around the network's predicted values. Our results show that DL techniques hold promise to infer physical parameters of Galactic rotational CCSN events.

\end{abstract}

\maketitle

\section{Introduction
\label{sec:intro}}

The gravitational collapse of the core of massive stars and the subsequent explosion as a supernova (a core-collapse supernova event, or CCSN event hereafter) is one of the most interesting sources of gravitational waves (GWs) to be detected in the coming years (see~\cite{Abdikamalov2022} for a recent review). CCSN events are thought to be the end result of stars with a zero age main sequence mass larger than about $8 M_{\odot}$. The core of these stars, consisting of iron-group nuclei, can no longer sustain nuclear burning and collapses. As the core reaches densities above nuclear saturation density and the equation of state (EOS) stiffens, the infalling material bounces back launching an outward-moving shock wave. Numerical simulations have long helped understand the processes by which the shock wave can gain sufficient energy to power a supernova explosion. Two main mechanisms have been proposed, neutrino-driven explosions and magneto-rotational explosions (see~\cite{Janka:2017,Abdikamalov2022} and references therein).  Rotation strongly affects the dynamics of CCSN events and, in turn, its GW emission. Most CCSN events are nonrotating or slowly rotating~\cite{Heger2005}. In these models, the GW emission is largely stochastic as it is triggered by convection and by the standing accretion shock instability.  In rapidly rotating stars, however, the early part of the GW signal is well determined. In these models the explosion is powered by the rotational kinetic energy  (see~\cite{Obergaulinger:2020} and references therein), leading to a bounce GW signal that is well understood. The GW strain of rapidly-rotating cores reaches its maximum right before bounce and its value depends on the corresponding degree of oblateness of the core. Numerical simulations have shown that for about 10 ms after core bounce the newly formed proto-neutron star (PNS) undergoes a series of oscillations driven by the excitation of axisymmetric fluid modes. Shortly after the GW signal becomes stochastic as a result of the growth of nonaxisymmetric hydrodynamical instabilities. One to three CCSN events are expected to occur in the Milky Way per century~\cite{Abbott:2020, Gossan2016} and less than 10\% are likely to be from fast-rotating progenitors~\cite{Raynaud2022}.

To date, all confident GW signals observed by the Advanced Laser Interferometer Gravitational Wave Observatory (LIGO) and Advanced Virgo detectors correspond to coalescing compact binaries (CBCs)~\cite{abbott_gwtc-1_2019, abbott_gwtc-2_2020, LIGOScientific:2021usb, GWTC-3}. The most recent optically-targeted searches for GWs from CCSN events, using data from the first three observing runs of LIGO and Virgo, found no evidence of significant candidates~\cite{Abbott:2020, Szczepanczyk2023}. As the sensitivity of the detectors increases the probability of detecting GW signals from CCSN events will also increase, especially within our Galaxy and nearby satellite galaxies~\cite{Abbott:2020, Abbott:2020_prospects, Szczepanczyk2021, Szczepanczyk2023, Mezzacappa2023, Lagos2023, Bruel:2023}. The successful detection of GW signals from CCSN events may help clarify the underlying physical processes occurring in the cores of massive stars during their gravitational collapse along with providing a brand new channel to infer properties of the source, allowing to probe the properties of the PNS, the nuclear EOS, the rotation of the core, and the explosion mechanism~\cite{Abdikamalov:2014, Powell2016,Richers,Morozova:2018,TF-universal-2019, Bizouard:2021, Afle:2021, Pajkos:2021, Sotani:2021, Andersen:2021,Saiz-Perez2022,Bruel:2023,Pastor-Marcos:2023,Powell2023,Wolfe:2023,Mitra:2023}. 

The detection of GW signals from CBCs is achieved through matched filtering~\cite{Abbott-guide}. This requires template waveform models to be precomputed with a faithful representation of true GWs. On the contrary, by their stochastic nature, CCSN GW signals are unmodeled, requiring a completely different approach for their detection. In this case, a coherent time-frequency analysis of the data in a network of detectors is used~\cite{cWB,Szczepanczyk2023}. Correspondingly, the estimation of the source parameters in the case of CBC signals is done through Bayesian inference. This method uses a large number of waveforms (or ``approximants'') that cover a wide parameter space and applies stochastic sampling techniques to evaluate their likelihood functions. 
Being computationally expensive, the Bayesian approach is, in general, not the optimal method to perform parameter inference of CCSN GW signals for several reasons. On the one hand, postbounce signals rapidly become stochastic a few milliseconds after core bounce. On the other hand, the amount of CCSN waveforms available is severely limited by the computational cost of the simulations and by the impossibility to build the waveforms using the same standard approximations employed for compact binaries. For the fairly brief ``deterministic'' part of the early bounce signal, matched-filtering analysis or Bayesian model selection can be used to infer source properties (see, e.g.,~\cite{Abdikamalov:2014, Pastor-Marcos:2023}). We also note that the dominant features of CCSN waveforms can be extracted with principal component analysis (PCA), where a mapping between their measured eigenvectors and the physical parameters of the progenitor star can be created for third-generation detectors~\cite{Afle:2021}. Moreover, for close enough sources, it is also possible to distinguish neutrino-driven CCSN events from magneto-rotational CCSN events using PCA~\cite{Rover2009, Logue2012, Powell2016, Powell2017, Powell2023}. 

An alternative to the methods mentioned before are machine learning (ML) techniques which have been shown to be significantly less time-consuming than Bayesian approaches in performing parameter estimation of CBC signals~\cite{Green:2020, dax_real-time_2021, williams_nested_2021, bayley_rapid_2022, Gabbard:2022, Dax:2023, bhardwaj2023peregrine}. 
The potential of ML for GW data analysis is becoming increasingly relevant, and the number and scope of applications are already important (see~\cite{Huerta:2019, Cuoco:2020, Zhao:2023, Stergioulas:2024} for recent reviews). Those include, e.g.,~detection methods for CBC signals~\cite{Osvaldo:2021, ALBUS, mockdatachal}, signal quality improvements~\cite{torres-forne_denoising_2016,torres-forne_application_2020}, waveform generation~\cite{Liao:2021,stefano_mlgw, stefano_bns}, and noise transient simulations in detectors~\cite{melissa_gans,lopez_gengli}. 
In the context of CCSN GW signals the application of ML techniques is also an active field of research~\cite{Astone:2018, Chan2020, Melissa:2021, Saiz-Perez2022,Iess:2023, Mitra:2023, Powell2023,  Sasaoka:2023}. The potential of convolutional neural networks (CNNs) to detect CCSN GW signals was first shown by~\cite{Astone:2018} using phenomenological waveforms injected in Gaussian noise. CNNs were also employed by~\cite{Chan2020} with 
CCSN GW signals injected in detector noise. At a false alarm probability of 10\%,  neutrino-driven explosions at 10~kpc yield an expected true alarm probability of 55\% (76\%)  for current (future) LIGO-Virgo-KAGRA (LVK) detectors, while for magneto-rotational explosions at 50~kpc the corresponding values increase to 84\% (92\%). Approaches based on learned dictionaries were first applied in~\cite{Saiz-Perez2022} for CCSN signal classification, obtaining 85\% true classifications on signals with a SNR from 15 to 20 (see also~\cite{Powell2023}). In~\cite{Melissa:2021} a mini-inception residual network (ResNet) for time-frequency images (also implementing convolutional layers) was used to improve the detectability of CCSN GW signals with the LVK pipeline coherent wave burst~\cite{cWB}. For a dataset comprising phenomenological CCSN waveforms in background noise from O2, an efficiency of around 80\%  was obtained for ${\rm SNR}\sim 16$. Recently~\cite{Mitra:2023} have used ML to investigate if it is possible to infer the iron core mass from the bounce and early ring-down GW signal of rapidly rotating CCSN models, and \cite{Iess:2023} have compared the performance of different CNNs and long short-term memory networks for multilabel classification of CCSN simulated signals and noise transients using real data. CNNs have also been used by~\cite{Sasaoka:2023} to classify GW from CCSN events using spectrograms from numerical simulations injected onto real noise data from O3. We also note the recent work by~\cite{Powell2023} which presents a comprehensive comparison of different methods (Bayesian model selection, dictionary learning, and CNNs) to determine the explosion mechanism from a GW CCSN detection using up-to-date waveforms from simulations (including, in particular, new three-dimensional (3D) long-duration magneto-rotational CCSN waveforms that cover the full explosion phase).

In this work, we present a deep learning (DL) approach for classification and parameter estimation using GWs from rapidly rotating CCSN events. We test residual CNN algorithms~\cite{resnet,rescnn} using datasets built from the numerically generated CCSN waveforms computed by~\cite{Richers}, which are injected into noise from the Advanced LIGO and Advanced Virgo detectors. 
We initially test these algorithms with spectrograms for classification as a proof of concept, building on previous group results on CBCs~\cite{Osvaldo:2021}. Additionally, we also test the same algorithms for classification and parameter estimation using time series. A different set of waveforms for rapidly rotating CCSN events, developed by~\cite{Mitra:2023}, is used to assess the generality of our approach. The tests in the time domain follow closely the recent work of~\cite{Pastor-Marcos:2023} and perform our inference on two key parameters, the GW strain amplitude ($\Delta h$) multiplied by the luminosity distance ($D$), $D \cdot \Delta h$, and the maximum (peak) frequency, $f_\textrm{peak}$, attained at core bounce. As discussed in~\cite{Pastor-Marcos:2023}, the results from~\cite{Richers}
for a 12~$M_\odot$ progenitor and a large group of equations of state show that when the ratio of rotational-kinetic energy to gravitational potential energy, $T/|W|$, is below 0.06 (i.e.~for slowly rotating cores), the first parameter is proportional to $T/|W|$, and the second one is proportional to the square root of the central density, $\sqrt{\rho_c}$. These relationships allow, in principle, the inference of PNS properties from the GW detection of a rotating CCSN event. In~\cite{Pastor-Marcos:2023} inference on these two parameters was conducted using Bayesian model selection employing a master waveform template built from the waveform catalog of~\cite{Richers}. Our approach, based on DL, can thus be regarded as complementary to the Bayesian approach of~\cite{Pastor-Marcos:2023}. As we show in this work, for rotating Galactic CCSN events, the inference of these two parameters using neural networks can be successfully achieved, yielding results within the $1\sigma$ band from the expected (true) values.


This paper is organized as follows: In Sec.~\ref{sec:waveforms} we describe the CCSN waveforms we select to train our models. In  Sec.~\ref{sec:GW_Spec} we discuss a first method in which the analysis of the signals is performed using spectrograms, focusing on signal classification only. Section~\ref{sec:GW_TS} presents a second, more complete method. Here, DL methods are applied to time series and we discuss results for both signal classification and parameter inference. Finally, our conclusions are outlined in Sec.~\ref{sec:conclusions}.

\section{Waveform selection \label{sec:waveforms}}

Our models are trained with waveforms from the Richers~\textit{et al.}~catalog~\cite{Richers}. 
This catalog is composed of 1824 numerical waveforms obtained from the collapse of a 12$M_\odot$ progenitor, using 18 different equations of state and covering a parameter space of 98 rotation profiles. The simulations from~\cite{Richers} focus on the bounce signal, including the collapse phase and the postbounce evolution up to $\sim 50$~ms after bounce. 
The bounce signal has been extensively studied in previous works~\cite{Zwerger1997, dimmelmeier2,Dimmelmeier2008, Ott2012, Abdikamalov:2014, Richers, Fuller2015, Edwards2021}, and it is considered the key signature for the presence of rotation in CCSN events. As mentioned before, our study will focus only on the part of the signal corresponding to the bounce and the following 10~ms.

The waveforms were obtained through a combination of 1D simulations of the collapse-phase deleptonization with the code \texttt{GR1D}~\cite{gr1d} and 2D core-collapse simulations with the \texttt{CoCoNuT Code}~\cite{coconut_paper}.
To produce the collapse-phase deleptonization, Richers \textit{et al.}~used an approximation proposed by Liebendörfer~\cite{liebendorfer} where the electron fraction ($Y_e$) is parametrized as a function of density ($\rho$) only from spherically symmetric (1D) nonrotating general relativity hydrodynamic simulations. This approximation was also used for the rotating case, as electron captures and neutrino interactions with matter are local and depend on the density in the collapse phase. In this case, the rotational flattening of the collapsing core can be considered relatively small.
%
The axisymmetric (2D) CCSN simulations were performed assuming the conformal flat condition approximation for the spacetime metric, and by forcing the initial model to rotate with constant angular velocity according to the rotation law 
\begin{equation}
    \Omega (r) = \Omega_0 \left[ 1 + \left( \frac{r}{A}\right)^2\right]^{-1},
    \label{rotation profile}
\end{equation}
\noindent
where $\Omega(r)$ is the angular velocity, $r$ is the distance from the axis of rotation, and $\Omega_0$ and $A$ are free parameters that determine the rotational speed/energy of the model and the distribution of angular momentum.

Following~\cite{Pastor-Marcos:2023}, from the Richers~\textit{et al.}~catalog~\cite{Richers} we only consider waveforms with $\Omega_0 \geq 3.0~{\textrm{rad}}{\textrm{s} }^{-1}$. This is done to avoid the numerical noise present in models with slow rotation rates, which gets amplified when performing the normalization of the signals. Moreover, we do not use any of the simulations that do not collapse within the first second (see Table~III of~\cite{Richers}). Considering the above constraints only 999 waveforms remain from the catalog. While the injections on the datasets used for the analyses with spectrograms employ the full catalog of 1824 waveforms, datasets created for the analyses with time series only apply the selected 999 waveforms. 

The waveforms assume an optimally oriented source. Therefore, for an axisymmetric system there is no cross-polarization ($h_\times$), and only the plus-polarization component ($h_+$) is different from zero, i.e.,~$h_\times = 0$ and $\quad h_+ = h_+^{\rm opt} \sin^2 \theta$.
Here, $\theta$ is the inclination angle between the rotational axis of the core and the line of sight of the observer. We fix this angle to $\pi/2$~rad, simplifying the expression above to $h_{+} = h_+^{\textrm{opt}}$. Each waveform on the catalog provides several physical parameters, from which we used the strain in units of distance ($D \cdot h_+^{\textrm{opt}}$), the timestamps $(t- t_{\textrm{b}})$, the peak frequency ($f_{\textrm{peak}}$), the maximum rotational velocity $\Omega_0$, and the GW strain amplitude ($\Delta h$) multiplied by the luminosity distance ($D$), ($D \cdot \Delta h$). 

\section{Analysis with Spectrograms \label{sec:GW_Spec}}

\subsection{Dataset generation and neural network details \label{sec:Sp_generation}}

\begin{figure}[t]
\centering
\includegraphics[width=0.49\textwidth]{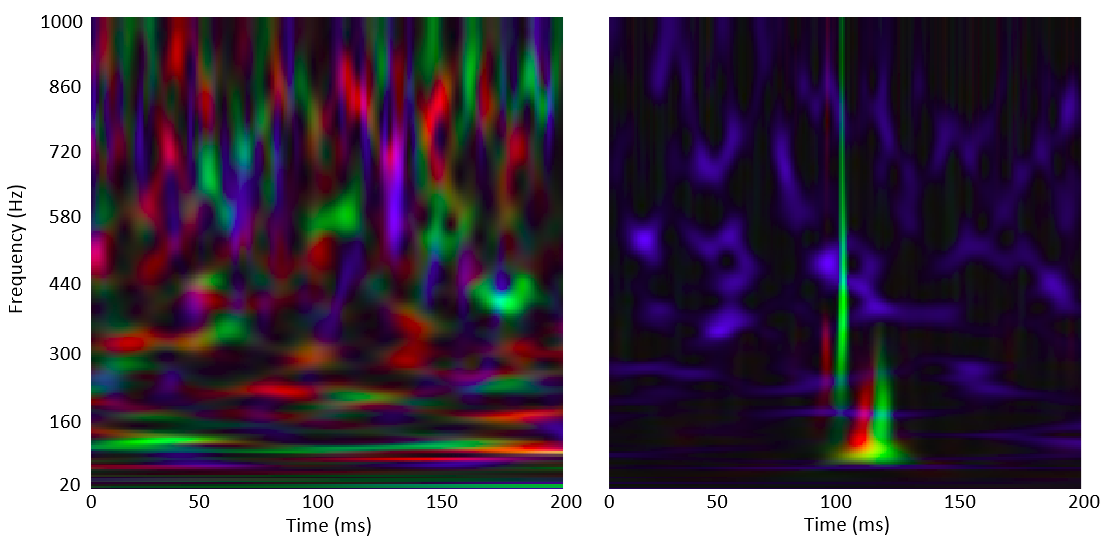}
\caption{\label{fig:spectrograms} RGB images of only real noise (left) and with an injected GW into real noise conditions (right) where the Hanford, Livingston, and Virgo interferometer spectrograms are represented in red, green, and blue, respectively.}
\end{figure}

Our first analysis of the waveforms is based on time-frequency representations of the signals or spectrograms. These diagrams represent the signal amplitude in pixel intensity, with time on the $x$ axis and frequency on the $y$ axis. The use of spectrograms is not arbitrary. Since the bounce and postbounce phases of CCSN events involve many frequencies, different patterns may be found in the time-frequency representation (produced by splitting the input time series into shorter sections and taking Fourier/wavelet transforms of these sections) when compared with the time-strain representation. These patterns, being contained in a short time interval and having many frequencies, are seen in a spectrogram as an intense narrow vertical band, as shown in Fig.~\ref{fig:spectrograms} (right).

In order to train and validate a neural network (NN) it is necessary to have a large enough data bank. To perform the classification test, we created a dataset consisting of images of only noise and others with injected signals. For both signal and noise spectrograms, we randomly selected time series segments of actual detector noise from O2 (corresponding to August 2017), with a duration of 4096~ms, without any specific criteria. We normalized the intensity of each frequency using a whitening procedure to account for variations in detector sensitivity across the frequency range. Our analysis considers frequencies between 20 and 1000~Hz, where the LIGO and Virgo detectors are most sensitive. We also notched out fixed frequencies affected by known artifacts, specifically 60, 120, and 240~Hz for the LIGO detectors and 50, 100, and 200~Hz for the Virgo detector, corresponding to the U.S. and European power grid frequency and their first harmonics, respectively. The sampling frequency was set to 16384~Hz. High-resolution time-frequency maps with a final time duration of 200~ms were constructed using the Q-transform method from \texttt{GWpy}~\cite{gwpy}, with its default parameters.
For the signal spectrograms, the only difference was the injection of a waveform at a random time position into a section of actual detector noise before whitening. We utilize all the 1824 waveforms from the Richers~\textit{et al.}~catalog~\cite{Richers}, described in Sec.~\ref{sec:waveforms}, assuming a constant sky position, fixing the distance to the source to 20~kpc and the right ascension, declination, and polarization angle to zero. 

The resulting images have dimensions of 256$\times$256 pixels, where the frequencies and time are distributed linearly on their respective axis. Since we use three detectors, the images are coded with RGB colors where red, green, and blue correspond to the spectrograms of LIGO-Hanford, LIGO-Livingston, and Virgo detectors, respectively. In addition, the intensity of each color varies between 0 and 255, so that when injecting a very intense signal, the normalization makes the noise almost imperceptible. In the left panel of Fig.~\ref{fig:spectrograms} a RGB image of only real background noise is shown, while the right panel displays a RGB image with an injected CCSN GW signal. 
In the latter, the signal stands out over the LIGO-Hanford and LIGO-Livingston noise. It shows two narrow stripes with a rapid variation in frequency, associated with the burst, and two subsequent features appearing shortly after, connected with the oscillations of the PNS. The noise is barely noticeable in the green and red channels, due to the normalization, while in the blue channel, the intensity of the injections is much lower than the noise (hence, the normalization makes the noise stand out).  Moreover, a delay can be observed between the detection of the two interferometers. 

A bank of $10^4$ images was generated, where half of them are only noise (labeled as ``background'') and the other half have CCSN GW signals injected into noise (labeled as ``signal''). This dataset was divided into two: a training set (85\%) and a validation set (15\%). In addition, the SNR of each signal was evaluated, serving as a filter to eliminate weak signals (SNR~<~5) that would otherwise only confuse our NN. 
We always consider the network SNR amplitude of the signals, which combines the individual SNR for each GW interferometer in the network, LIGO-Hanford, LIGO-Livingston, and Virgo, defined as
\begin{equation}
    \textrm{network SNR} = \sqrt{\sum_{i=1}^3 {\textrm{SNR}_i}^2}.
\end{equation}
The individual $\textrm{SNR}_i$ is found by calculating the cross-correlation matrix between the postinjected strain and the injected template after projecting into the corresponding detector. Both strains are treated with whitening, bandpass, and notch filtering, as explained before. The maximum value of this cross-correlation matrix is considered as the $\textrm{SNR}_i$.

As our study is based on images, we work with CNNs. These must be capable of differentiating between background and signal. We use a pretrained ResNet101 model which implements an architecture with a residual CNN 101 layers deep~\cite{deep_paper_7780459}. We load the ResNet101 model pretrained with over one million images from the ImageNet database~\cite{ImageNet}. The pretrained network can classify images into $1000$ object categories. As a result, the network has learned feature-rich representations for a wide range of images, although we will only need it to distinguish between two classes, background and signal. To accomplish this, we need to adapt the network by changing its last layer of 1000 neurons to one with only 2 of them. For the generation of the datasets, functions from the \texttt{PYTHON}~\cite{python} libraries \texttt{SciPy}~\cite{scipy}, \texttt{PyCBC}~\cite{pycbc}, and \texttt{GWpy}~\cite{gwpy} were used. We also used the library \texttt{fast.ai}~\cite{fastai} for the classification tests.

\subsection{Classification results with spectrograms \label{sec:class_Sp}}

\begin{figure}[t]
\centering
\includegraphics[width=0.35\textwidth]{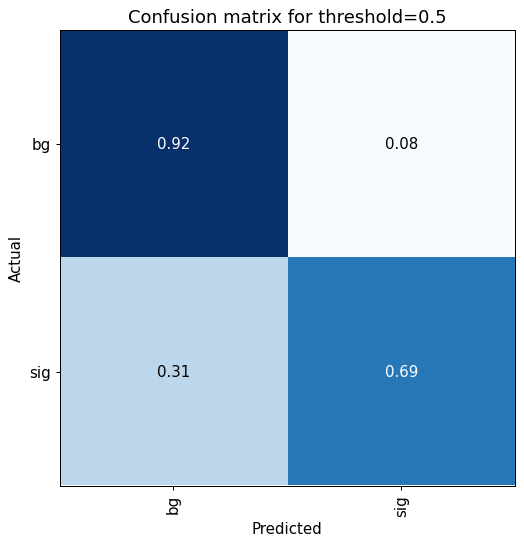}
\caption{\label{fig:confmatrixSP} Confusion matrix for the classification test with the 13th epoch weights showing the fractions of strains predicted as background noise or signal versus their actual labels.}
\end{figure}

Our classification test aims at distinguishing the dataset elements where a CCSN GW signal has been injected from those where only detector background noise is present. For a training of 25 epochs, the network found the minimum error in the 13th epoch. The model obtained had an accuracy\footnote{The accuracy is defined as $\frac{TP + TN}{TP + TN + FP + FN}$, where $TP$ ($TN$) corresponds to the correctly classified signals (background) and $FP$ ($FN$) corresponds to the backgrounds (signals) that are misclassified.} of 82\%, providing the confusion matrix displayed in Fig.~\ref{fig:confmatrixSP}. This figure shows that 92\% of the background images are correctly classified, with a false positive rate (FPR) of 8\%. While the model only classifies 69\% of the signals correctly, a precision of 90\% is obtained. These results show that our model detects signals with high confidence, even if it only detects a modest amount of GW signals.

\subsection{Testing the classification model}

\begin{figure}[t]
\begin{center}
\subfloat[]{%
\includegraphics[scale=0.2]{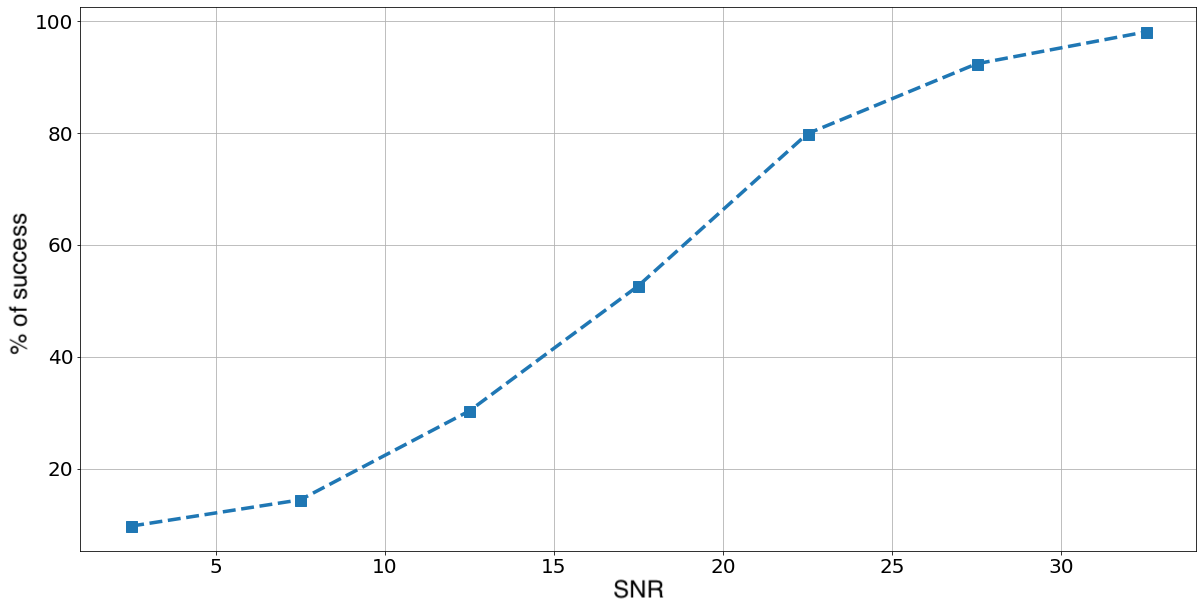}%
}
\\
\subfloat[]{%
\hspace*{-1mm}\includegraphics[scale=0.2]{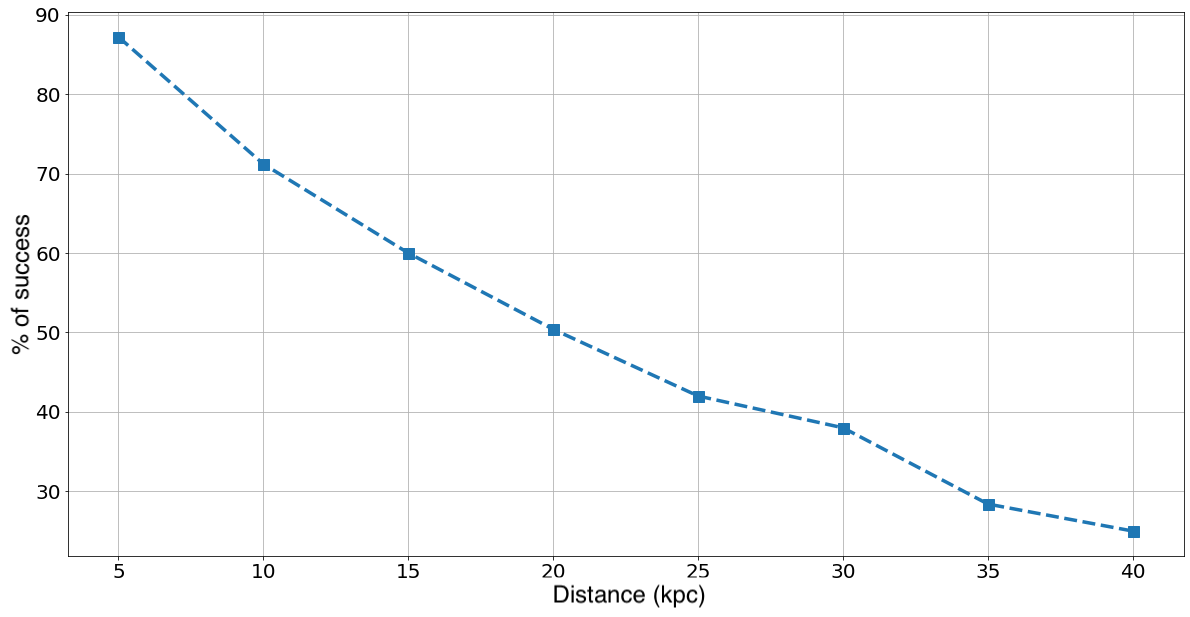}%
}
\caption{Percentage of success in the validation of our network to detect a CCSN GW signal as a function of SNR (a) and distance (b).}
\label{fig:succeses_SNR_distances}
\end{center}
\end{figure}

We further tested our model by performing additional studies with two different datasets sorted by SNR and by distance to the source. One dataset consists of 1500 images arranged in groups differentiated by five SNR units, with fixed distance and sky position, similar to the datasets used for training. The other dataset consists of 4000 images with distances ranging between 5 and 40~kpc, keeping a fixed sky position and without any limits on the SNR. Figure~\ref{fig:succeses_SNR_distances}(a) shows the relationship between the percentage of success of the network in detecting signals and the SNR of the injected GW. It is interesting to highlight that an accuracy of 95\% and a precision of 93\% is reached for SNR~>~30. 
Although the network was trained to detect a GW generated at a distance of 20~kpc, it can also be used to infer GW signals at different distances. 
In Fig.~\ref{fig:succeses_SNR_distances}~(b), we observe that at shorter distances, the percentage of success is higher, as expected, decreasing as the source gets farther away.
We observe a discrepancy for a distance of 20~kpc with respect to the value obtained previously during validation. This is due to the fact that when generating images for training and validation, those with SNR~<~5 were eliminated, as mentioned above, a condition that we did not apply here.

\section{Analysis with Time Series \label{sec:GW_TS}}

\subsection{Dataset generation \label{sec:TS_generation}}

In our second method, we apply DL techniques in the time domain. While significant patterns may be more easily recognized in the time-frequency representation, some information might be lost during the Q-transform process, which would not occur when using time series directly. In this case, we only use the 999 selected waveforms from the Richers~\textit{et al.}~catalog, described in Sec.~\ref{sec:waveforms}. For each of the selected waveforms, we scale the strain $D \cdot h_+^{\textrm{opt}}$ to a randomly chosen distance of the source between 5 and 20~kpc (i.e.,~a CCSN that occurs in our Galaxy). For the background noise we use publicly available data from both LIGO (Hanford and Livingston) and Virgo, with an initial Global Positioning System (GPS) time $t_{\textrm{GPS}}=1253326755~\textrm{s}$, which corresponds to the end of September 2019, i.e.,~around the middle of the O3 run. We define a random time $t_0 \in [6, 1750]~\textrm{s}$ and select a segment of the noise time series with a window $[t_0 - 5, t_0 + 10]~\textrm{s}$ for the background sample. The waveforms were down-sampled to match the noise sampling frequency. In this analysis we used a sampling frequency of 4096~Hz, allowing faster performance without losing information. Each waveform was projected into each detector with random values for the angles defining the sky position (declination, right ascension, and polarization angle). Then, the projected waveforms are injected into noise samples for each detector, having the time of bounce of the signal ($t_{\textrm{b}}$) randomly set in the interval $[t_0, t_0 + 0.80]~\textrm{s}$, for the specific case of datasets with one-second time window. For completeness, we also consider datasets of 0.50~s and 0.25~s with $t_{\textrm{b}}$ in the range $[t_0, t_0 + 0.40]~\textrm{s}$ and $[t_0, t_0 + 0.20]~\textrm{s}$, respectively.

We perform whitening on the resulting strains (both for the background and signal classes) through inverse spectrum truncation using the amplitude spectral density of the three detectors' noise. We then apply a bandpass filter from 20 to 1000~Hz and notch filters at the individual frequencies of 60, 120, and 240~Hz for the two LIGO strains and 50, 100, and 200~Hz for the Virgo strains. We finally apply a crop on the strains to get the expected time window (1.00, 0.50, and 0.25~s). At this point, a selection is made on the resulting whitened background section to exclude high-intensity glitches (such as blips). After whitening, the maximum absolute amplitude is around $1.5\times 10^{-21}$, so we only allow segments with a maximum absolute amplitude below $2.7\times 10^{-21}$. 

\begin{figure}[t]
\begin{center}
\subfloat[]{%
\includegraphics[clip,scale=0.6]{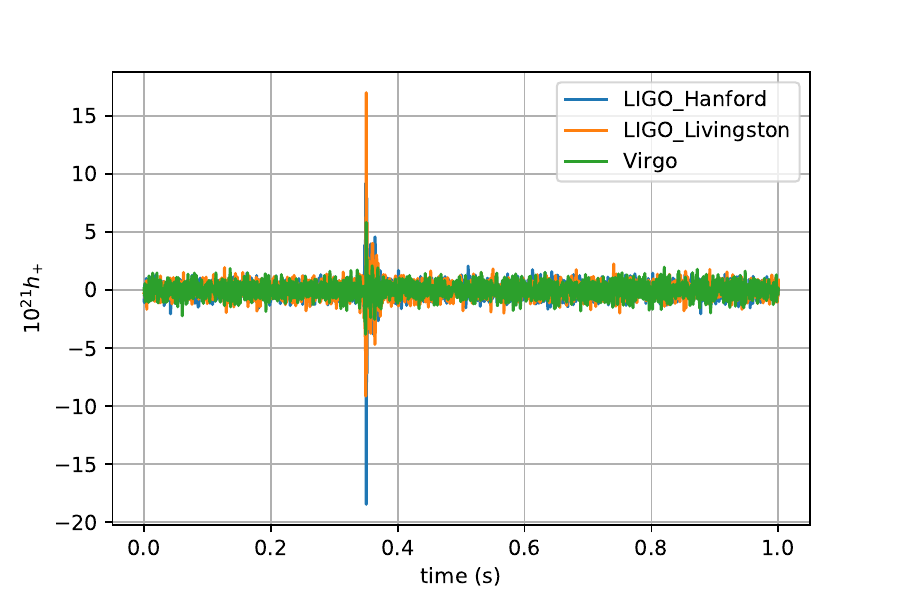}%
}
\\
\subfloat[]{%
\includegraphics[clip,scale=0.6]{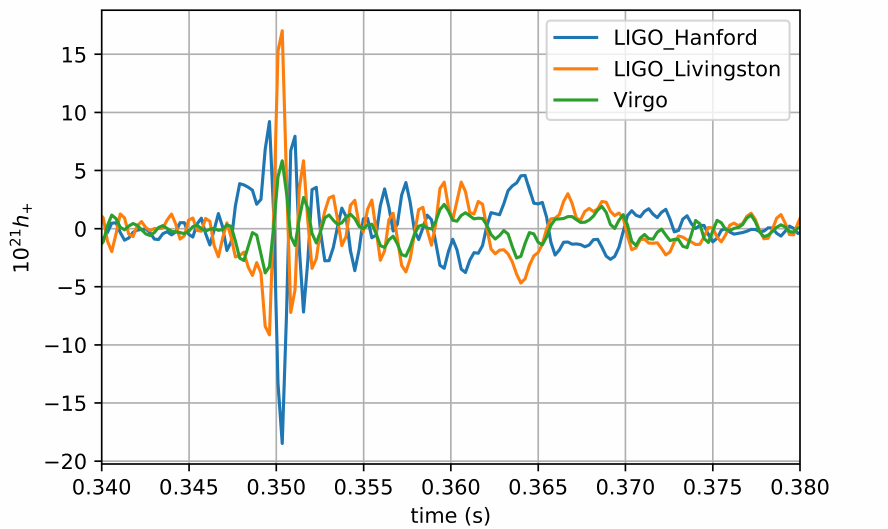}%
}
\caption{Example of a dataset's element labeled as signal, with a waveform injected into noise from O3a, with a time window of 1~s (a) and 0.04~s (b).}
\label{fig:sig}
\end{center}
\end{figure}

\begin{figure}[t]
\begin{center}
\subfloat[]{%
\hspace*{-1mm}\includegraphics[clip,scale=0.6]{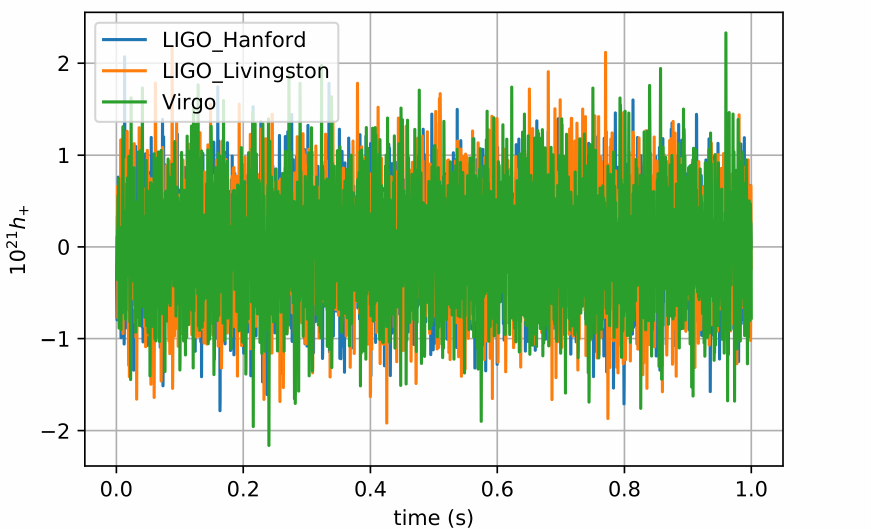}%
}
\\
\subfloat[]{%
\hspace*{-1mm}\includegraphics[clip,scale=0.6]{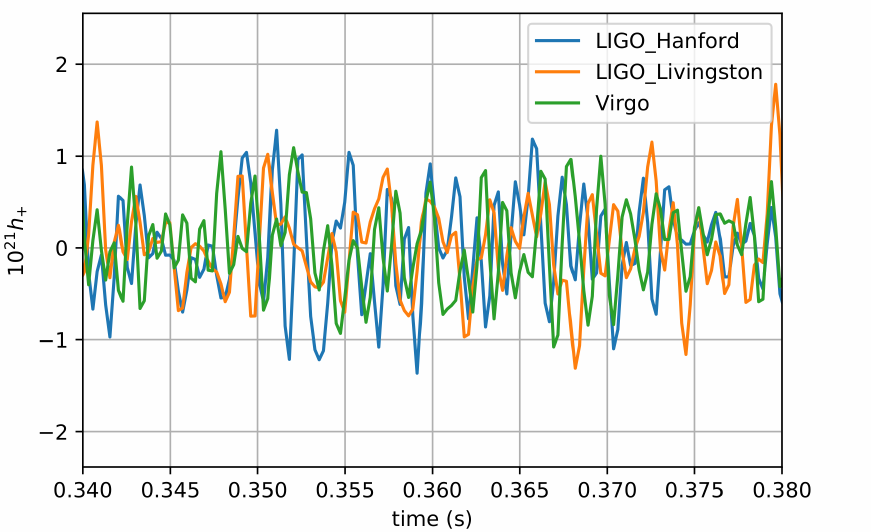}%
}
\caption{Example of a dataset's element labeled as background, with only noise from O3a, with a time window of 1~s (a) and 0.04~s (b).}
\label{fig:bg}
\end{center}
\end{figure}

For the classification and parameter estimation tests in the time domain, we use datasets of $10^4$ elements, where each element is composed of three time series, one for each detector. Examples of these time series are given in Figs.~\ref{fig:sig} and~\ref{fig:bg}, corresponding to one element of the signal and background samples, respectively. Table~\ref{tab:datasets} characterizes the datasets used for classification and parameter estimation with time series. For the generation of the datasets, functions from the \texttt{PYTHON}~\cite{python} libraries \texttt{SciPy}~\cite{scipy}, \texttt{PyCBC}~\cite{pycbc}, and \texttt{GWpy}~\cite{gwpy} were used. We also used the libraries \texttt{FAST.AI}~\cite{fastai} and \texttt{TSAI}~\cite{tsai} for the classification and parameter estimation tests, described in the next sections.

\subsection{Deep neural network: Architecture and methodology \label{sec:TS_DL_network}}

\begin{figure}[ht]
\centering
\includegraphics[width=0.5\textwidth]{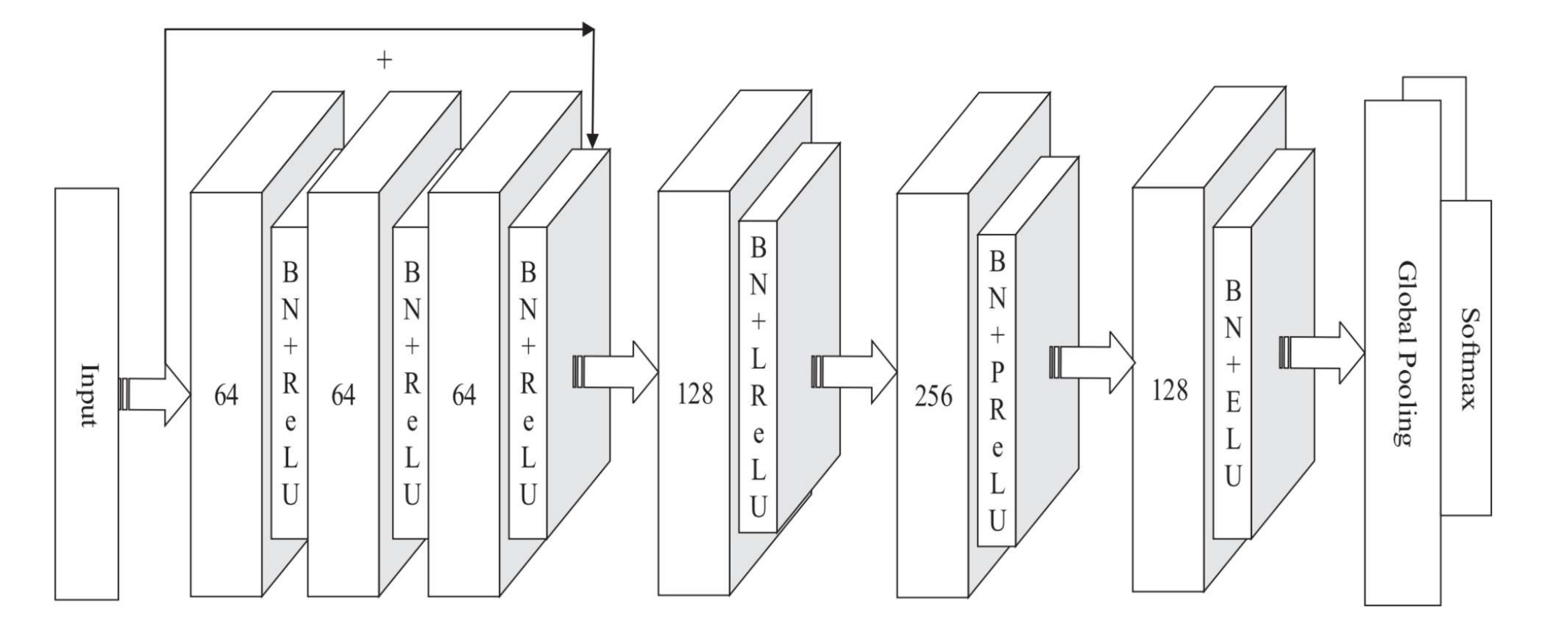}
\caption{\label{fig:rescnn} Representation of the ResCNN's architecture as presented in~\cite{rescnn}.}
\end{figure}

Time-domain GW signals of CCSN events are tested, both for classification and parameter estimation, using a residual convolutional neural network (ResCNN). This network is an implementation of the CNN with residual learning blocks, proposed in~\cite{rescnn}. It has six convolutional layers and uses batch normalization techniques and different activation functions. The most used activation function is the rectified linear unit (ReLU). Other versions of this activation function are also used in the architecture, i.e., leaky ReLU (LReLU), parametric ReLU (PReLU), and exponential linear unit (ELU). It is desirable to use different activation functions in different layers to achieve better performance~\cite{rescnn}.
As the analysis of the time series needs the precise location of the signal features, the pooling operation was removed. Another reason for this choice is the fact that pooling reduces the number of inputs given to the next layer, limiting the information available.
For this architecture, the usual fully connected layer at the end of the network was replaced by a global average pooling and a final softmax layer, so that the output of the network is a probability distribution of its own predictions.
The architecture of this network is represented in Fig.~\ref{fig:rescnn}. 

\begin{table*}[t]
\setlength\doublerulesep{4mm} 
\setlength\tabcolsep{4pt}
\begin{tabular}{|c|c|c|}
\toprule
\hline
Parameters & Classification & ~ Parameter estimation ~ \\
\hline
\makecell{~ \\ $\textrm{SNR} > 5$ \\ $ \textrm{progenitor's mass} = 12 \ M_\odot$ ~\\
         ~$\textrm{distance} \in [5, 20]$~kpc ~ \\
         $ \textrm{inclination} =\frac{\pi}{2}$ ~\\~
         $ \textrm{declination} \in [-\pi, \pi]$ ~\\~
         $ \textrm{right ascension} \in [0, 2\pi]$ ~ \\~
         $ \textrm{polarization} \in [0, 2\pi]$ \\~} & \makecell{Training set: 80\% \\
                                            ~Validation set: 20\% ~} 
                                & \makecell{Training set: 70\% \\
                                            ~Validation set: 30\% ~} 
                                \\
                                
\midrule
\hline
\end{tabular}
\caption{Description of the datasets used to perform classification using time series and parameter estimation.}
\label{tab:datasets}
\end{table*}

We use the function \texttt{fit$\_$one$\_$cycle} to perform the training of the network, defining the maximum learning rate (LR) and the weight decay. This function of \texttt{FAST.AI} internally uses the usual \texttt{fit} function but applies the 1cycle policy for better results~\cite{smith17}. 
The maximum LR and weight decay used should be adapted to each dataset and architecture~\cite{smith18}. In our case, to find the best values for our datasets we used another \texttt{FAST.AI} function called \texttt{lr$\_$find} based on the LR range test~\cite{smith15}.
We verified that for our datasets the best results obtained were for weight decays and maximum LR of the order of $10^{-3}$.



For our parameter estimation computations, before giving any dataset to the network, we perform a standard normalization of the signals' parameters. To find the predictions of the network in both classification and parameter estimation, we use Monte Carlo (MC) dropout~\cite{mcdropout}, passing each signal through the network 100 times. The prediction of the network is the mean value of all the 100 evaluations. Finally, we apply \texttt{PYTHON'S} PCA function from \texttt{SCIKIT-LEARN}~\cite{sklearn} on the two-dimensional distributions of predicted versus true values, obtained on the parameter estimation tests, according to what follows. Although PCA is typically used for dimensionality reduction by identifying the most relevant features in datasets, we do not use this capability of PCA here. Instead, we use the \texttt{PYTHON} function solely to perform singular value decomposition (SVD) on the calibration plots, comparing predicted values to true values of $D \cdot \Delta h$ and $f_\textrm{peak}$. We extract the two orthogonal eigenvectors to calculate the slope, and with the singular values
we obtain the standard deviation  of the distributions, $\sigma$.

\subsection{Classification results with time series \label{sec:TS_class}}

\begin{figure}[t]
\centering
\includegraphics[width=0.4\textwidth]{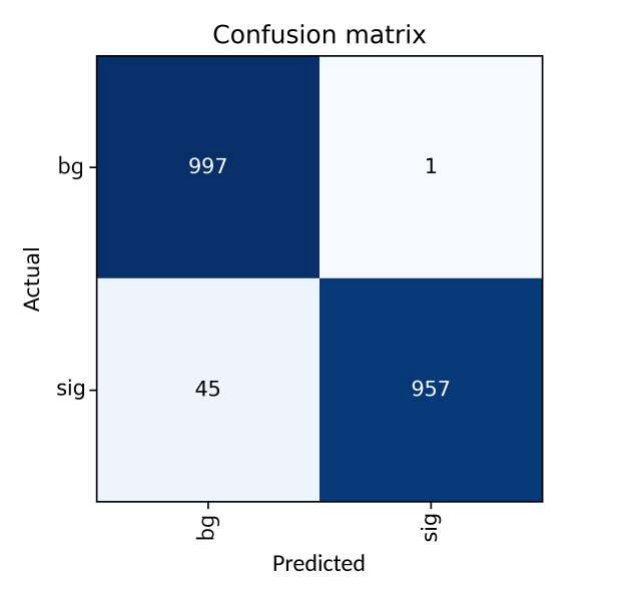}
\caption{\label{fig:matrix12} Confusion matrix for the classification test with 12 epochs giving the number of strains predicted as background or signal versus their assigned labels.}
\end{figure}

\begin{figure}[ht]
\centering
\includegraphics[width=0.5\textwidth]{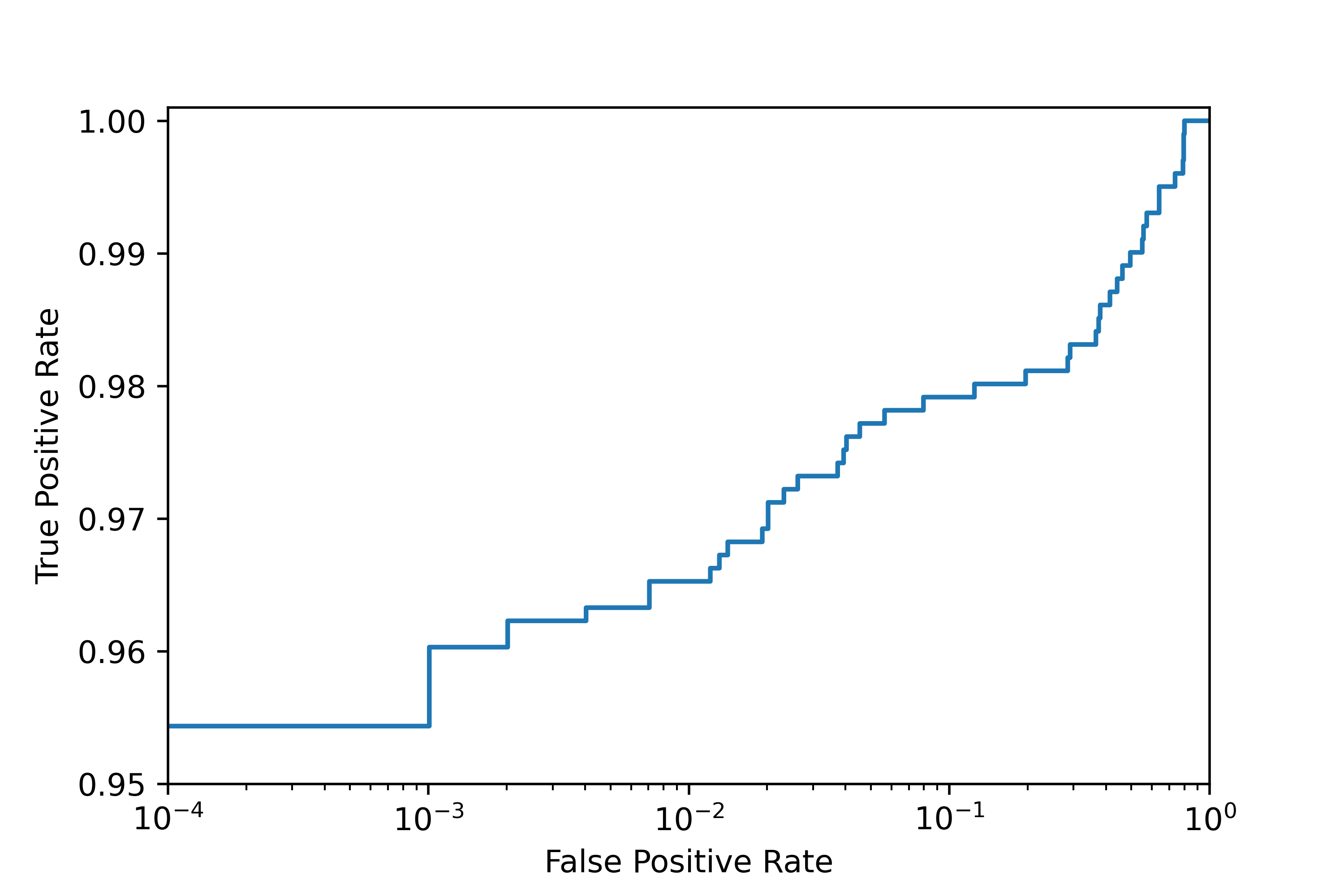}
\caption{\label{fig:roc12} ROC curve of the classification test with 12 epochs.}
\end{figure}

A classification test aims at obtaining a model that best allows to distinguish elements of a dataset where a CCSN GW signal has been injected from those of only background noise. Our best results are achieved for a 12-epoch training and considering a 0.50 threshold, which led to a 0.09 minimum validation loss. This model has an accuracy of 98\%, providing the confusion matrix shown in Fig.~\ref{fig:matrix12}. Only 45 actual signals were predicted as background, which corresponds to 4.5\% of all GW signals present in the validation set. In addition, only one element labeled as background was misclassified as a signal, giving a FPR of 0.10\%. Figure~\ref{fig:roc12} displays the receiver operating characteristic (ROC) curve of this classification model, with a 0.99 area under the curve.

\begin{figure}[ht]
\includegraphics[width=0.5\textwidth]{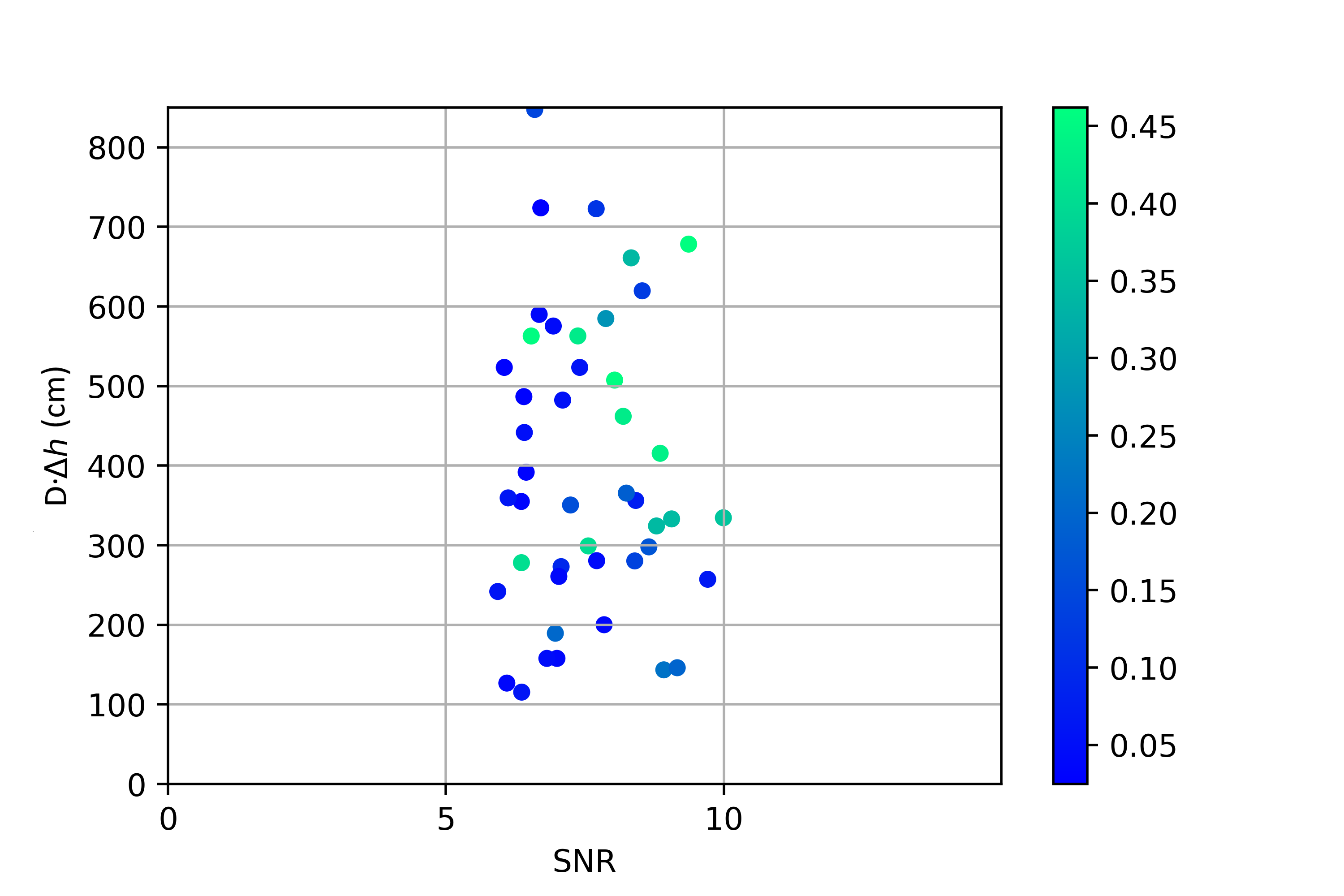}
\includegraphics[width=0.5\textwidth]{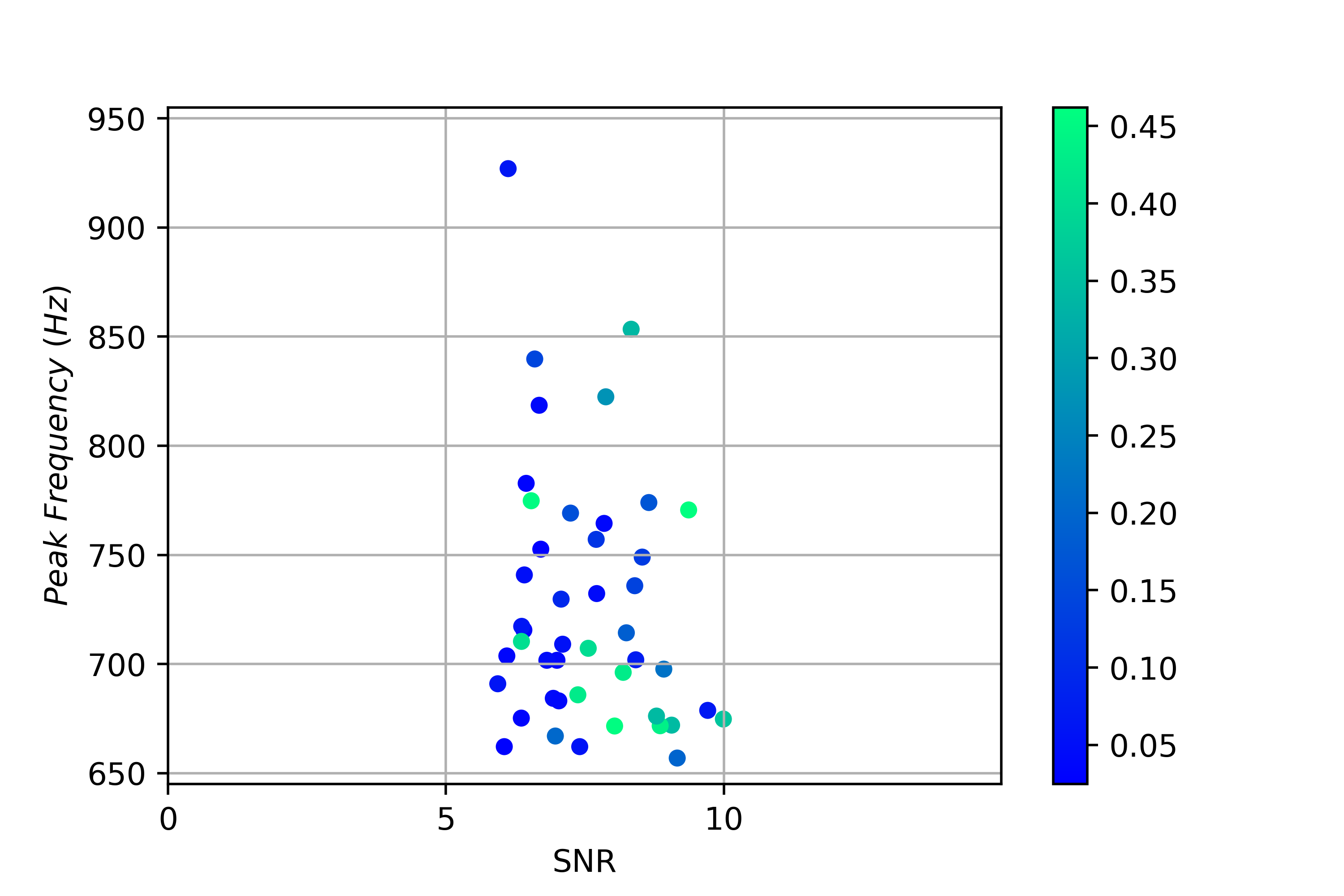}
\centering
\caption{\label{fig:Dh12_fpeak12} Values of $D \cdot \Delta h$ (top panel) and $f_\textrm{peak}$ (bottom panel) as a function of the SNR for the 45 misclassified signals on the validation set. The color bar represents the score given by the model to each element.}
\end{figure}

To better understand the results of the misclassified 45 signals (false negatives), Fig.~\ref{fig:Dh12_fpeak12} shows the distribution of the model's scores for each of those signals. In this plot we show our two key CCSN parameters, $D \cdot \Delta h$ (top panel) and $f_\textrm{peak}$ (bottom panel), as a function of the SNR. The different colors of the points represent the prediction scores given by the model. All misclassified signals have SNR~$\le$~10, which is expected since in those cases it is difficult to distinguish the signal from the random peaks of the noise.

For completeness we have built a spectrograms dataset corresponding to the O3a data used here, to test the network applied in Sec.~\ref{sec:class_Sp} to spectrograms. As expected, when changing the background noise of the spectrograms to O3a, the accuracy of the network goes up to 95\%, reflecting the better noise conditions of O3a when compared to O2.

\subsection{Parameter inference results with time series \label{sec:TS_parameter_inf}}

A parameter inference test was performed to predict the values of the GW strain amplitude ($\Delta h$) multiplied by the luminosity distance ($D$), $D \cdot \Delta h$ (the difference between the highest and lowest points in the bounce signal normalized to the distance) and the peak frequency, $f_\textrm{peak}$ (the highest frequency measured in the first 6~ms after bounce). As discussed above (see also~\cite{Pastor-Marcos:2023}) these two parameters are the most relevant ones to characterize the waveform at bounce when $T/|W|<0.06$, as $D \cdot \Delta h \propto T/|W|$ and $f_\textrm{peak}\propto \sqrt{\rho_c}$ for a large group of equations of state~\cite{Richers}.

Three datasets of $10^4$ elements each were analyzed for different choices of time windows, i.e.~1.00, 0.50, and 0.25~s, as mentioned previously. For all the parameter estimation tests, we trained the network for 30 epochs setting the LR to $2\times 10^{-3}$ and the weight decay to $10^{-3}$. As expected, there is a relation between the size of the time window used and the training time needed for each epoch. For the 1.00~s dataset, each epoch took around 90~s to complete. By decreasing the window to 0.50~s, the time needed for each epoch also decreased to around 50~s. And lastly, for the 0.25~s dataset, each epoch took around 27~s to complete. As we decrease the time window of the dataset, the training time also decreases, almost in a linear way.

\begin{figure}[t]
\begin{center}{%
\includegraphics[clip,scale=0.063]{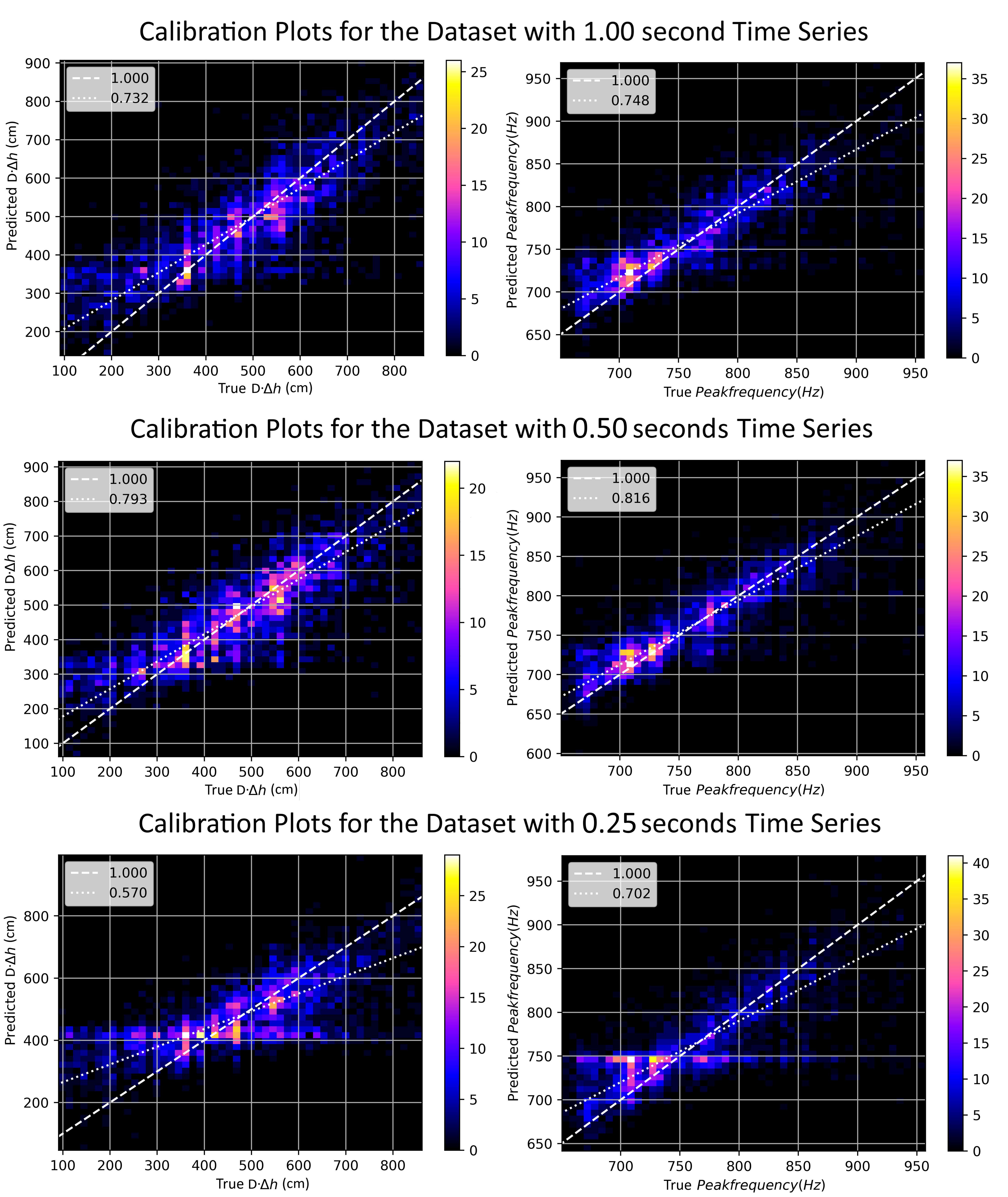}%
}
\caption{Calibration plots of predicted vs true $D \cdot \Delta h$ (left) and predicted vs true $f_\textrm{peak}$ (right) with SNR $\geq 5$ for the different datasets (corresponding to different time windows). The white dashed line represents the ideal case, where the prediction value is equal to the true value (slope is 1), and the white dotted line represents the slope of the distribution calculated with the PCA function. The specific values of the slopes are indicated in the legends.}
\label{fig:reg_cuts_Dh5}
\end{center}
\end{figure}

\begin{figure}[t]
\begin{center}{%
\includegraphics[clip,scale=0.063]{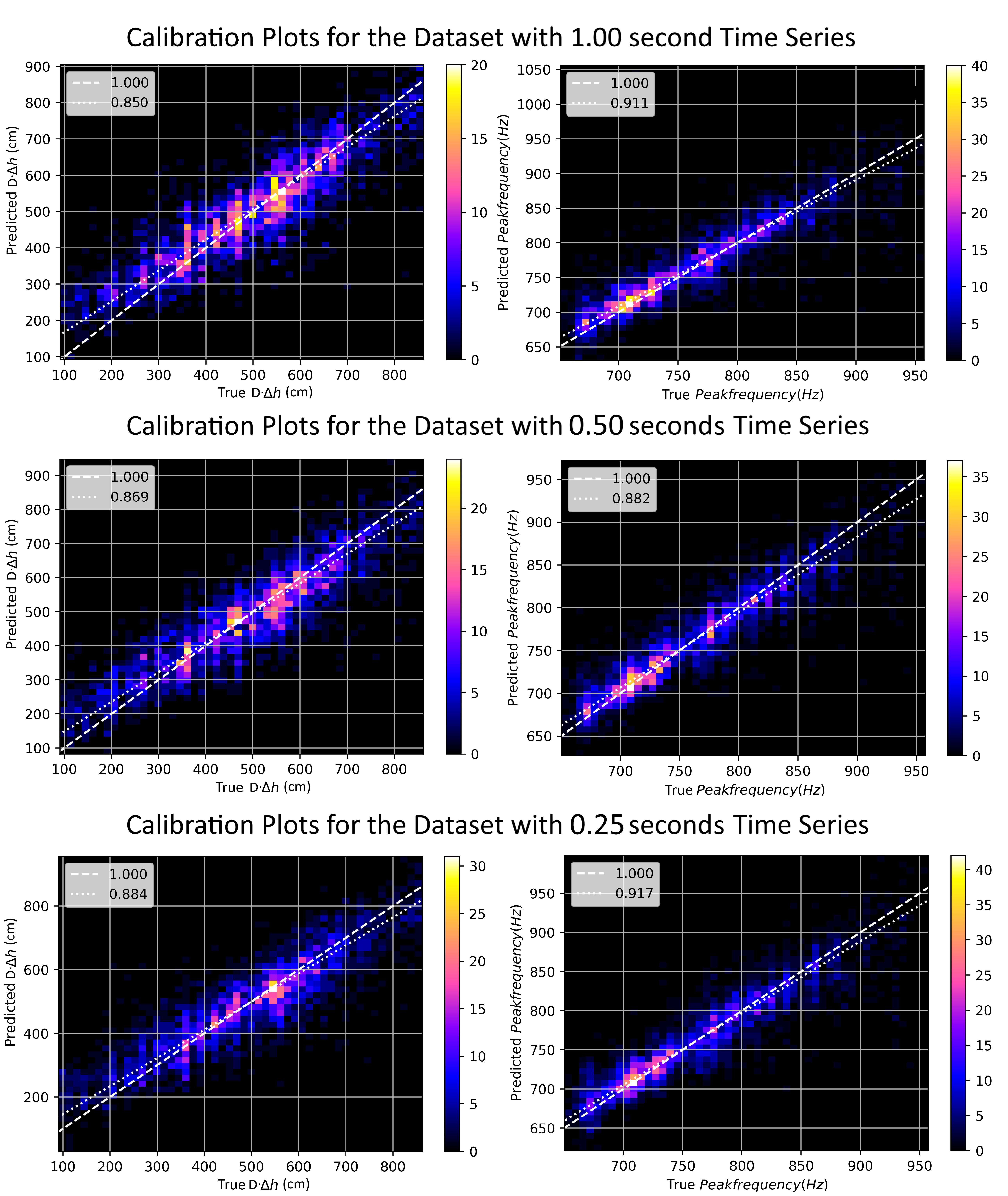}
}
\caption{Calibration plots of predicted vs true $D \cdot \Delta h$ (left) and predicted vs true $f_\textrm{peak}$ (right) with SNR $\geq 15$ for the different datasets. As in Fig.~\ref{fig:reg_cuts_Dh5} the white dashed line represents the ideal case (slope 1), and the white dotted line represents the slope of the distribution calculated with the PCA function.}
\label{fig:reg_cuts_Dh15}
\end{center}
\end{figure}

Figure~\ref{fig:reg_cuts_Dh5} shows the calibration plots for the three datasets, considering SNR~$\geq$~5. These plots represent predicted values versus true values for $D \cdot  \Delta h$ (left) and $f_\textrm{peak}$ (right). Two white lines are present in each plot. The dotted lines represent the slope of each distribution, which should be close to the ideal slope, 1.00, represented by the white dashed line.
The best results are obtained for the 0.50~s dataset (middle panel in the figure).
Comparing the calibration plots for the two parameters, our models yield more precise predictions for $f_\textrm{peak}$ than for $D \cdot \Delta h$. Moreover, low values of $D \cdot \Delta h$ and $f_\textrm{peak}$ generally tend to be overestimated.

It is worth noticing that horizontal scattered lines are visible in Fig.~\ref{fig:reg_cuts_Dh5} for all the distributions of the calibration plots. In order to understand this feature we generated new datasets with a minimum SNR of 15, consistent with other studies which place constraints on the minimum SNR estimated to detect GW signals from CCSN events~\cite{Szczepanczyk2021}. The new results for the parameter estimation of $D \cdot \Delta h$ and $f_\textrm{peak}$ on the three different datasets are shown in Fig.~\ref{fig:reg_cuts_Dh15}. 
With this choice of minimum SNR the horizontal scattered lines disappear. We also see an improvement in the values of the slopes, with the best results corresponding to the 0.25~seconds dataset. In this case, for the $D \cdot \Delta h$ distribution, the slope is $0.884 \pm 0.008$, and the standard deviation is $\sigma_{D \cdot \Delta h} = (52.6 \pm 10.7)$~cm. For the $f_\textrm{peak}$ distribution, the results yield a slope of $0.917 \pm 0.006$ and a standard deviation of $\sigma_{f_\textrm{peak}} = (18.3 \pm 3.9)$~Hz.

\begin{figure}[t]
\centering
\includegraphics[width=0.5\textwidth]{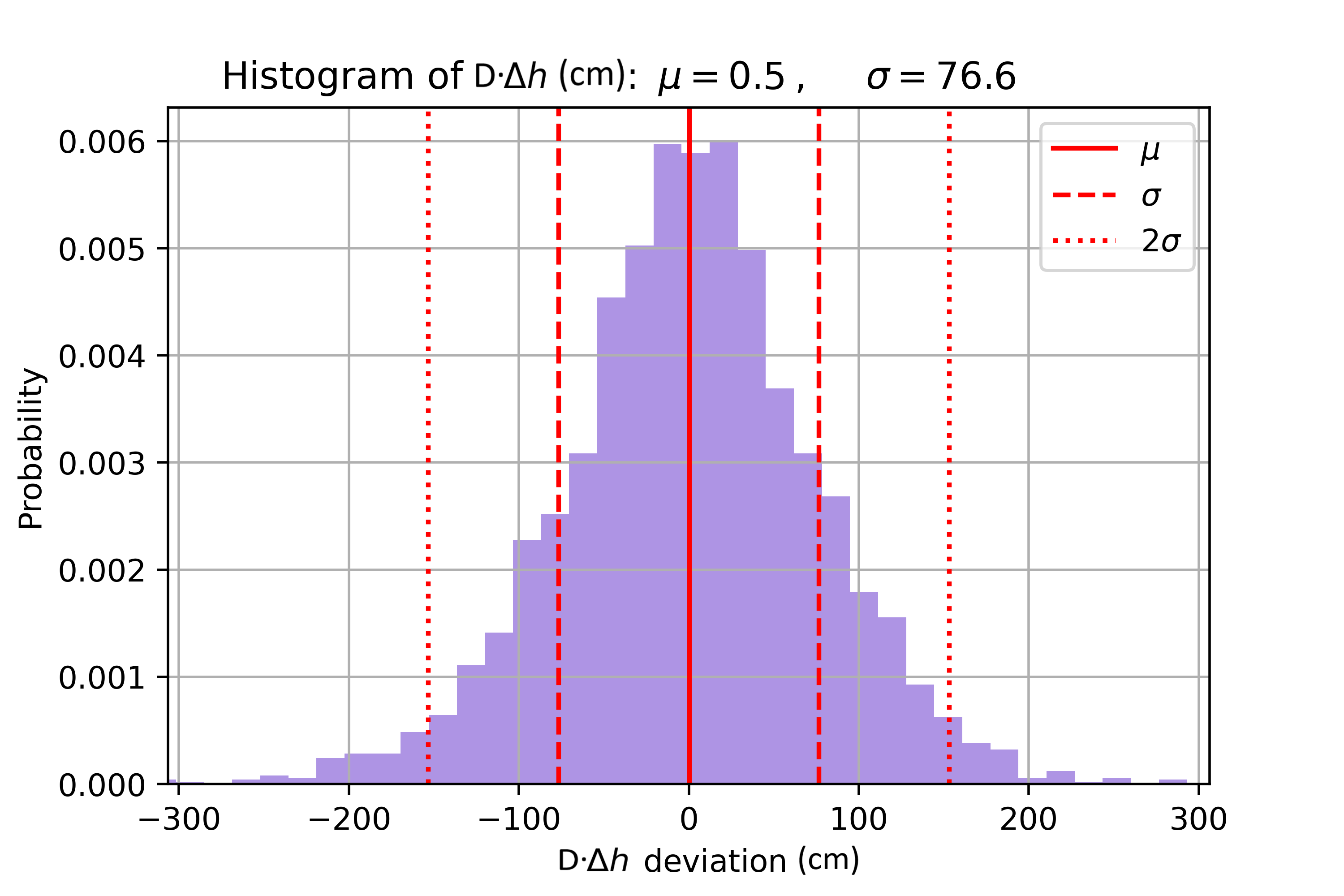}
\includegraphics[width=0.5\textwidth]{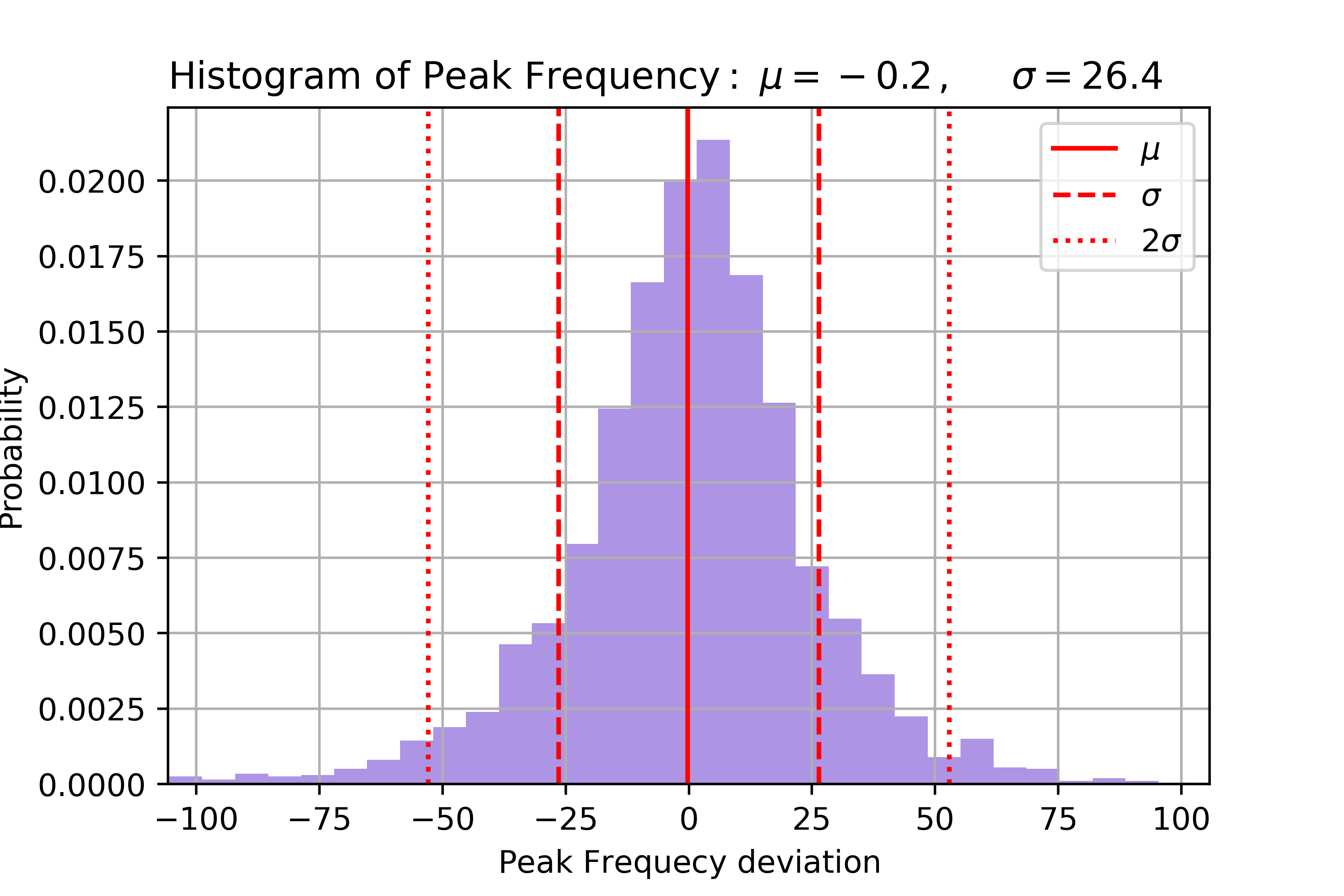}
\caption{\label{fig:prob_Dh_fpeak} Probability distribution of the deviations of $D \cdot \Delta h$ (top panel) and $f_\textrm{peak}$ (bottom panel), i.e., the difference between the labeled
value and the predicted value, for the 0.25~seconds dataset and a minimum SNR of 15.}
\end{figure}

The probability distributions of the difference between true and predicted values of $D \cdot \Delta h$ and $f_\textrm{peak}$ are shown in the two plots of Fig.~\ref{fig:prob_Dh_fpeak} for the 0.25~seconds dataset with minimum SNR of 15. For $D \cdot \Delta h$, a mean value of $\mu_{[D \cdot \Delta h_\textrm{true} - D \cdot \Delta h_\textrm{pred.}]}=0.5~\pm~76.6$~cm is obtained. As for the $f_\textrm{peak}$ case, the mean value obtained is $\mu_{[f_\textrm{peak, true} - f_\textrm{peak, pred.}]} =~-0.2~\pm~26.4$~(Hz). We can clearly see that the inference of $f_\textrm{peak}$ is much more accurate than $D \cdot \Delta h$. Taking into account that in our datasets $D \cdot \Delta h \in [100, 850]$~cm and $f_\textrm{peak} \in [650, 950]$~Hz, we can consider the mean values and standard deviations obtained quite reasonable.

\subsection{Testing the network with a different catalog \label{sec:new_catalog}}

We now discuss the performance of our network with a different set of CCSN signals.
In the absence of an actual discovery of a CCSN GW signal to test our best model, we resort to employing waveforms from a different catalog within the same parameter space of the datasets used for training. The new waveforms were selected from the catalog created recently by Mitra~$\textit{et al.}$~\cite{Mitra:2023, different_catalog}. This catalog contains 404 waveforms of progenitors with masses ranging from 12 to 40$M_\odot$, albeit only for a single EOS, SFHo~\cite{SF}. While training our network with the Richers {\it et al.}~catalog~\cite{Richers} we only considered 12$M_\odot$ progenitors. Therefore, we do the same for this additional test. The CCSN simulations of~\cite{Mitra:2023, different_catalog} also employ the general relativistic hydrodynamics code \texttt{CoCoNuT}~\cite{coconut_paper}, as in  Richers' simulations. While the simulations extend up to 25~ms after bounce, we only consider the part of the signals in the first 10~seconds after bounce in order to keep consistency between training and testing.


\begin{table*}[!htpb]
\begin{tabular}{c|c|c|c|c|c|c|c|c|c}
\hline
Model A1 & & & & & & & & & \\
\hline
\hline
Waveform & \texttt{s12A1005.5} & \texttt{s12A1006} & \texttt{s12A1006.5} & \texttt{s12A1007} & \texttt{s12A1007.5} & \texttt{s12A1008} & \texttt{s12A1008.5} & \texttt{s12A1009} &  \\
$D \cdot \Delta h$ (cm) & 422 $\pm$ 50 & 446 $\pm$ 54 & 501 $\pm$ 59 & 591 $\pm$ 76 & 676 $\pm$ 57 & 730 $\pm$ 49 & 770 $\pm$ 45 & 757 $\pm$ 45 & \\
 & [\textbf{374}] & [\textbf{449}] & [\textbf{533}] & [\textbf{611}] & [\textbf{678}] & [\textbf{747}] & [\textbf{799}] & [{832}] & -- \\
$f_\textrm{peak}$ (Hz) & 755 $\pm$ 12 & 752 $\pm$ 12 & 762 $\pm$ 12 & 790 $\pm$ 25 & 827 $\pm$ 26 & 848 $\pm$ 16 & 883 $\pm$ 17 & 892 $\pm$ 30 & \\
& [{770}] & [\textbf{764}] & [{783}] & [{826}] & [\textbf{843}] & [\textbf{863}] & [\textbf{986}] & [{923}] & \\
\hline
Model A3 & & & & & & & & & \\
\hline
\hline
Waveform & \texttt{s12A3003.5} & \texttt{s12A3004} & \texttt{s12A3004.5} & \texttt{s12A3005} & \texttt{s12A3005.5} & \texttt{s12A3006} & \texttt{s12A3006.5} & \texttt{s12A3007} & \texttt{s12A3007.5} \\
$D \cdot \Delta h$ (cm) & 449 $\pm$ 58 & 493 $\pm$ 63 & 543 $\pm$ 61 & 647 $\pm$ 56 & 648 $\pm$ 39 & 637 $\pm$ 36 & 596 $\pm$ 31 & 561 $\pm$ 29 & 512 $\pm$ 36 \\
& [\textbf{361}] & [\textbf{464}] & [\textbf{564}] & [\textbf{643}] & [\textbf{666}] & [\textbf{629}] & [\textbf{583}] & [\textbf{560}] & [\textbf{520}] \\
$f_\textrm{peak}$ (Hz) & 765 $\pm$ 11 & 764 $\pm$ 13 & 763 $\pm$ 18 & 787 $\pm$ 19 & 808 $\pm$ 17 & 831 $\pm$ 15 & 846 $\pm$ 13 & 851 $\pm$ 14 & 831 $\pm$ 19 \\
& [\textbf{772}] & [\textbf{773}] & [\textbf{780}] & [\textbf{798}] & [\textbf{816}] & [\textbf{841}] & [\textbf{858}] & [\textbf{862}] & [{859}] \\
\hline
Model A4 & & & & & & & & & \\
\hline
\hline
Waveform & \texttt{s12A4003} & \texttt{s12A4003.5} & \texttt{s12A4004} & \texttt{s12A4004.5} & \texttt{s12A4005} & \texttt{s12A4005.5} & \texttt{s12A4006} & \texttt{s12A4006.5} &  \\
$D \cdot \Delta h$ (cm) & 445 $\pm$ 59 & 484 $\pm$ 58 & 538 $\pm$ 61 & 597 $\pm$ 52 & 570 $\pm$ 33 & 520 $\pm$ 35 & 477 $\pm$ 28 & 442 $\pm$ 22 &  \\
& [{354}] & [\textbf{462}] & [\textbf{551}] & [\textbf{589}] & [\textbf{559}] & [\textbf{506}] & [\textbf{468}] & [\textbf{440}] &  -- \\
$f_\textrm{peak}$ (Hz) & 761 $\pm$ 12 & 759 $\pm$ 12 & 759 $\pm$ 15 & 775 $\pm$ 14 & 790 $\pm$ 16 & 811 $\pm$ 14 & 808 $\pm$ 12 & 798 $\pm$ 15 & \\
& [\textbf{768}] & [\textbf{765}] & [\textbf{770}] & [\textbf{786}] & [\textbf{802}] & [\textbf{816}] & [\textbf{820}] & [{817}] & \\
\hline
Model A5 & & & & & & & & & \\
\hline
\hline
Waveform & \texttt{s12A5003} & \texttt{s12A5003.5} & \texttt{s12A5004} & \texttt{s12A5004.5} & \texttt{s12A5005} &  & &  &  \\
$D \cdot \Delta h$ (cm) & 463 $\pm$ 62 & 506 $\pm$ 68 & 550 $\pm$ 61 & 560 $\pm$ 42 & 532 $\pm$ 38 & & & & \\
& [{387}] & [{494}] & [\textbf{552}] & [\textbf{544}] & [\textbf{506}] & -- & -- & -- & -- \\
$f_\textrm{peak}$ (Hz) & 759 $\pm$ 12 & 760 $\pm$ 16 & 766 $\pm$ 14 & 780 $\pm$ 14 & 801 $\pm$ 15 & & & & \\
& [\textbf{764}] & [\textbf{764}] & [\textbf{775}] & [{797}] & [\textbf{813}] &  &  &  & \\
\hline
\end{tabular}
\caption{Parameter estimation results for the 30 selected waveforms from the Mitra~$\textit{et al.}$~catalog~\cite{Mitra:2023, different_catalog}. The models are grouped by the degree of differential rotation $A$, corresponding from top to bottom to 300~km (A1), 634~km (A3), 1268~km (A4), and 10000~km (A5). For each waveform we report the \textbf{prediction}, for both $D \cdot \Delta h$ and $f_{\textrm{peak}}$, in the form $\mu \pm \sigma$ and the true value in brackets. When the true values are within $1\sigma$ of the predictions they are written in boldface. }
\label{tab:results_new_catalog}
\end{table*}

We selected 30 waveforms of this new catalog and, for each one of them, generated a dataset with 100 elements as we did before. We fixed the time window to 0.25~s and the minimum SNR to 15, which corresponds to the conditions that led to our best model. Moreover, we allowed the model to vary its response by using MC dropout. By performing signal injections in different noise conditions as well as changes in the model response, we obtain a more realistic implementation of a true GW detection.
For a given dataset, prediction values were determined for each element and, with the collection of the 100 predicted values, the mean value $\mu$ and the standard deviation $\sigma$ were calculated for both parameters, $D \cdot \Delta h$ and $f_\textrm{peak}$. 

The results of the parameter estimation test with this new catalog are reported in Table~\ref{tab:results_new_catalog}. The name of the waveforms in this table encodes the characteristics of the signal: \texttt{s12} signifies that the mass of the progenitor star is 12$M_\odot$; \texttt{A1}, \texttt{A3}, \texttt{A4}, and \texttt{A5} correspond to the degree of differential rotation of the precollapse star [see Eq.~(\ref{rotation profile})] of 300, 634, 1268, and 10000~km, respectively, and the remainder of the string name gives the central angular velocity in $\textrm{rad} \, \textrm{s}^{-1}$.
Most of our predictions are within 1$\sigma$ from the true values, confirming that our model would be able to infer physical parameters of a Galactic CCSN GW signal in more realistic conditions. For $D \cdot \Delta h$ only four waveforms are outside of this interval but all are covered by the 2$\sigma$ band. In the case of the $f_\textrm{peak}$ predictions, only seven waveforms have true values outside of the 1$\sigma$ interval, but all are within 2$\sigma$.

\section{Conclusions} \label{sec:conclusions}

The first detection of a GW signal from a CCSN is highly anticipated as it can provide important information about the physical processes occurring during the gravitational collapse of massive stars. In this paper we have discussed a method to extract information through the analysis of simulated CCSN waveforms. In particular, we have developed DL techniques to perform classification and parameter inference of rapidly rotating CCSN events using the information encoded in the early (nonstochastic) part of their gravitational waveforms.
To do so we have trained NNs on a catalog of numerically generated signals~\cite{Richers} injected into real LIGO-Virgo background noise from O2 and O3a, both in the form of spectrograms (time-frequency diagrams) and time series. The results presented in this work provide further support to the usefulness of DL methods for GW data analysis. In the specific context of CCSN, the methods discussed here seem worthy of being implemented in CCSN detection (classification) and parameter estimation pipelines.

Following previous work on CBC GW signals~\cite{Osvaldo:2021}, we have first presented a classification test using a ResNet101 on a dataset composed of $10^4$ spectrograms. CCSN GW signals were injected into real noise from O2 of the LIGO Hanford, LIGO Livingston, and Virgo detectors making up half of the dataset. All signals have a fixed distance of 20~kpc, a fixed sky position, and a duration of 200~ms. The other half of the dataset contained only background noise spectrograms, which had to be separated by the network's model. The model obtained after training accomplished an accuracy of 82\% and a precision of 90\%. We have also analyzed the influence of SNR and distance to the source on the network's performance. For SNR~>~30 the accuracy of the model becomes 95\%. In addition, we attain more than 50\% of success on the classification only for distances below 20~kpc. While there is room for improvement, we see that using spectrograms in DL shows great promise, with results aligning with the values achieved in~\cite{Osvaldo:2021} for CBCs in O2 noise. The percentage of false positives (and false negatives) is increased due to the numerous glitches in O2 and the lower sensitivity of the detectors at that time. This issue can be minimized using background noise from more recent observing runs. As expected, when changing the background noise of the spectrograms to O3a, the accuracy of the network goes up to 95\%, reflecting the better noise conditions of O3a when compared to O2.

For the second and main part of our study, we have considered time series. This classification test was performed with a ResCNN on a dataset composed of $10^4$ time series where CCSN GW signals had been injected into 1-second-long sections of real noise from O3a of the two LIGO detectors and the Virgo detector. For this method, we have variable distance, from 5 to 20~kpc, and random sky position. All the tests performed with time series have been done with residual CNNs applied to datasets composed of numerically generated signals from different waveforms. The classification test was accomplished with a $10^4$ dataset consisting of equal parts of elements with only background noise from O3a and elements of injected signals. The network has provided a model with an accuracy of 98\% with only one false positive. These results confirm the efficiency of these networks regarding classification and show that, even with a small dataset, high values of accuracy can be obtained. By further inspecting the false negatives we noticed a correlation with the SNR of the signals, as all of the misclassifications occurred for elements with SNR~<~10. These results show the excellent performance of deep learning techniques for detection, with the time series method reflecting slightly better results than the values achieved with spectrograms in O3a.

To perform parameter estimation we have used three datasets composed only of injected signals from the catalog of~\cite{Richers}. These datasets differ in their time duration: 1.00, 0.50, and 0.25~s.
Considering a minimum SNR value of 15, the best results are obtained with the 0.25 s dataset. By performing SVD, we have found that the slope of the distribution of the $D \cdot \Delta h$ calibration plot is $0.884 \pm 0.008$ and the standard deviation $\sigma_{D \cdot \Delta h, 15} = (52.6 \pm 10.7)$~cm. For the peak frequency, the value of the slope obtained is $0.917 \pm 0.006$ with $\sigma_{f_\textrm{peak}, 15} =( 18.3 \pm 3.9)$~Hz standard deviation. As a final test, we have assessed our best model with 30 additional waveforms from a new catalog of rapidly rotating CCSN models, computed recently by~\cite{Mitra:2023}. For each waveform, we have created a dataset composed of 100 elements with 0.25~s duration and SNR~$\geq$~15. The results obtained have shown that most of the true values are within 1$\sigma$ of the prediction distribution, confirming that our model might be able to infer parameters of fast-rotating Galactic CCSN GW signals.

While the results reported in this work show the potential of deep learning for inferring properties of rotational CCSN events, they can nonetheless be improved. An obvious improvement would be accomplished by enlarging the datasets. The datasets used for the training of our models only have $10^{4}$ elements, which is considered a small size for the kind of classification and regression tests with neural networks. However, enlarging the datasets would require having available more waveforms from CCSN simulations, which are computationally costly. We note that we are currently working on the computation of new fast-rotating models for different progenitor masses and expect to present an updated study of our findings elsewhere. In addition, in this study, we have used background noise from the existing network of GW detectors (namely Advanced LIGO and Advanced Virgo). As current detectors are improved and third-generation detectors become operational, the sensitivity will improve and the background noise will be reduced, which will lead to higher SNR values and easier recognition of CCSN GW signals.

\section*{Acknowledgements}

We thank Pablo Cerd\'a-Dur\'an, Anastasios Theodoropoulos, José D. Martín, Nino Villanueva, and Pedro Passos for fruitful discussions during the course of this work. G. E. acknowledges the Spanish MINECO Grants No. MINECO/FEDER, Projects No. PID2021-122547NB-I00 FIS2021, MCIN with funding from European Union NextGenerationEU PR47/21 MADQuantum-CM PRTR-CM, and Ministry of Economic Affairs Quantum ENIA Project.
O. G. F. is supported by an FCT doctoral scholarship (reference UI/BD/154358/2022). J. A. F. and A. T.-F. are supported by the Spanish Agencia Estatal de Investigación (grant PID2021-125485NB-C21) funded by MCIN/AEI/10.13039/501100011033 and ERDF A way of making Europe. Further support is provided by the Generalitat Valenciana (Grant No. CIPROM/2022/49), by the EU's Horizon 2020 research and innovation (RISE) Programme H2020-MSCA-RISE-2017 (FunFICO-777740), and by the European Horizon Europe staff exchange (SE) Programme HORIZON-MSCA-2021-SE-01 (NewFunFiCo-101086251). A. O. in partially supported by FCT, under the contract CERN/FIS-PAR/0037/2021. This material is based upon work supported by NSF's LIGO Laboratory which is a major facility fully funded by the National Science Foundation.

\bibliographystyle{apsrev}
\bibliography{references}

\begin{thebibliography}{89}
\expandafter\ifx\csname natexlab\endcsname\relax\def\natexlab#1{#1}\fi
\expandafter\ifx\csname bibnamefont\endcsname\relax
  \def\bibnamefont#1{#1}\fi
\expandafter\ifx\csname bibfnamefont\endcsname\relax
  \def\bibfnamefont#1{#1}\fi
\expandafter\ifx\csname citenamefont\endcsname\relax
  \def\citenamefont#1{#1}\fi
\expandafter\ifx\csname url\endcsname\relax
  \def\url#1{\texttt{#1}}\fi
\expandafter\ifx\csname urlprefix\endcsname\relax\def\urlprefix{URL }\fi
\providecommand{\bibinfo}[2]{#2}
\providecommand{\eprint}[2][]{\url{#2}}

\bibitem[{\citenamefont{Abdikamalov et~al.}(2022)\citenamefont{Abdikamalov, Pagliaroli, and Radice}}]{Abdikamalov2022}
\bibinfo{author}{\bibfnamefont{E.}~\bibnamefont{Abdikamalov}}, \bibinfo{author}{\bibfnamefont{G.}~\bibnamefont{Pagliaroli}}, \bibnamefont{and} \bibinfo{author}{\bibfnamefont{D.}~\bibnamefont{Radice}}, \emph{\bibinfo{title}{Gravitational Waves from Core-Collapse Supernovae}} (\bibinfo{publisher}{Springer Nature Singapore}, \bibinfo{address}{Singapore}, \bibinfo{year}{2022}), ISBN \bibinfo{isbn}{978-981-16-4306-4}, \urlprefix\url{https://doi.org/10.1007/978-981-16-4306-4_21}.

\bibitem[{\citenamefont{{Janka}}(2017)}]{Janka:2017}
\bibinfo{author}{\bibfnamefont{H.-T.} \bibnamefont{{Janka}}}, in \emph{\bibinfo{booktitle}{Handbook of Supernovae}}, edited by \bibinfo{editor}{\bibfnamefont{A.~W.} \bibnamefont{{Alsabti}}} \bibnamefont{and} \bibinfo{editor}{\bibfnamefont{P.}~\bibnamefont{{Murdin}}} (\bibinfo{year}{2017}), p. \bibinfo{pages}{1095}.

\bibitem[{\citenamefont{{Heger} et~al.}(2005)\citenamefont{{Heger}, {Woosley}, and {Spruit}}}]{Heger2005}
\bibinfo{author}{\bibfnamefont{A.}~\bibnamefont{{Heger}}}, \bibinfo{author}{\bibfnamefont{S.~E.} \bibnamefont{{Woosley}}}, \bibnamefont{and} \bibinfo{author}{\bibfnamefont{H.~C.} \bibnamefont{{Spruit}}}, \bibinfo{journal}{Astrophys. J.} \textbf{\bibinfo{volume}{626}}, \bibinfo{pages}{350} (\bibinfo{year}{2005}), \eprint{astro-ph/0409422}.

\bibitem[{\citenamefont{{Obergaulinger} and {Aloy}}(2020)}]{Obergaulinger:2020}
\bibinfo{author}{\bibfnamefont{M.}~\bibnamefont{{Obergaulinger}}} \bibnamefont{and} \bibinfo{author}{\bibfnamefont{M.~{\'A}.} \bibnamefont{{Aloy}}}, \bibinfo{journal}{Mon. Not. R. Astron. Soc.} \textbf{\bibinfo{volume}{492}}, \bibinfo{pages}{4613} (\bibinfo{year}{2020}), \eprint{1909.01105}.

\bibitem[{\citenamefont{{Abbott} et~al.}(2020{\natexlab{a}})\citenamefont{{Abbott}, {Abbott}, {Abbott}, {Abraham}, {Acernese}, {LIGO Scientific Collaboration}, and {Virgo Collaboration}}}]{Abbott:2020}
\bibinfo{author}{\bibfnamefont{B.~P.} \bibnamefont{{Abbott}}}, \bibinfo{author}{\bibfnamefont{R.}~\bibnamefont{{Abbott}}}, \bibinfo{author}{\bibfnamefont{T.~D.} \bibnamefont{{Abbott}}}, \bibinfo{author}{\bibfnamefont{S.}~\bibnamefont{{Abraham}}}, \bibinfo{author}{\bibfnamefont{F.}~\bibnamefont{{Acernese}}}, \bibinfo{author}{\bibnamefont{{LIGO Scientific Collaboration}}}, \bibnamefont{and} \bibinfo{author}{\bibnamefont{{Virgo Collaboration}}}, \bibinfo{journal}{Phys. Rev. D} \textbf{\bibinfo{volume}{101}}, \bibinfo{eid}{084002} (\bibinfo{year}{2020}{\natexlab{a}}), \eprint{1908.03584}.

\bibitem[{\citenamefont{{Gossan} et~al.}(2016)\citenamefont{{Gossan}, {Sutton}, {Stuver}, {Zanolin}, {Gill}, and {Ott}}}]{Gossan2016}
\bibinfo{author}{\bibfnamefont{S.~E.} \bibnamefont{{Gossan}}}, \bibinfo{author}{\bibfnamefont{P.}~\bibnamefont{{Sutton}}}, \bibinfo{author}{\bibfnamefont{A.}~\bibnamefont{{Stuver}}}, \bibinfo{author}{\bibfnamefont{M.}~\bibnamefont{{Zanolin}}}, \bibinfo{author}{\bibfnamefont{K.}~\bibnamefont{{Gill}}}, \bibnamefont{and} \bibinfo{author}{\bibfnamefont{C.~D.} \bibnamefont{{Ott}}}, \bibinfo{journal}{\prd} \textbf{\bibinfo{volume}{93}}, \bibinfo{eid}{042002} (\bibinfo{year}{2016}), \eprint{1511.02836}.

\bibitem[{\citenamefont{{Raynaud} et~al.}(2022)\citenamefont{{Raynaud}, {Cerd{\'a}-Dur{\'a}n}, and {Guilet}}}]{Raynaud2022}
\bibinfo{author}{\bibfnamefont{R.}~\bibnamefont{{Raynaud}}}, \bibinfo{author}{\bibfnamefont{P.}~\bibnamefont{{Cerd{\'a}-Dur{\'a}n}}}, \bibnamefont{and} \bibinfo{author}{\bibfnamefont{J.}~\bibnamefont{{Guilet}}}, \bibinfo{journal}{Mon. Not. R. Astron. Soc.} \textbf{\bibinfo{volume}{509}}, \bibinfo{pages}{3410} (\bibinfo{year}{2022}), \eprint{2103.12445}.

\bibitem[{\citenamefont{Abbott et~al.}(2019)}]{abbott_gwtc-1_2019}
\bibinfo{author}{\bibfnamefont{B.~P.} \bibnamefont{Abbott}} \bibnamefont{et~al.}, \bibinfo{journal}{Phys. Rev. X} \textbf{\bibinfo{volume}{9}}, \bibinfo{pages}{031040} (\bibinfo{year}{2019}), \bibinfo{note}{1811.12907}.

\bibitem[{\citenamefont{Abbott et~al.}(2021{\natexlab{a}})}]{abbott_gwtc-2_2020}
\bibinfo{author}{\bibfnamefont{R.}~\bibnamefont{Abbott}} \bibnamefont{et~al.}, \bibinfo{journal}{Phys. Rev. X} \textbf{\bibinfo{volume}{11}}, \bibinfo{pages}{021053} (\bibinfo{year}{2021}{\natexlab{a}}), \eprint{2010.14527}.

\bibitem[{\citenamefont{Abbott et~al.}(2021{\natexlab{b}})}]{LIGOScientific:2021usb}
\bibinfo{author}{\bibfnamefont{R.}~\bibnamefont{Abbott}} \bibnamefont{et~al.} (\bibinfo{year}{2021}{\natexlab{b}}), \eprint{2108.01045}.

\bibitem[{\citenamefont{{The LIGO Scientific Collaboration} et~al.}(2021)\citenamefont{{The LIGO Scientific Collaboration}, {the Virgo Collaboration}, and {the KAGRA Collaboration}}}]{GWTC-3}
\bibinfo{author}{\bibnamefont{{The LIGO Scientific Collaboration}}}, \bibinfo{author}{\bibnamefont{{the Virgo Collaboration}}}, \bibnamefont{and} \bibinfo{author}{\bibnamefont{{the KAGRA Collaboration}}}, \bibinfo{journal}{arXiv}  (\bibinfo{year}{2021}), \urlprefix\url{https://arxiv.org/abs/2111.03606#}.

\bibitem[{\citenamefont{{Szczepa{\'n}czyk} et~al.}(2023)\citenamefont{{Szczepa{\'n}czyk}, {Zheng}, {Antelis}, {Benjamin}, {Bizouard}, {Casallas-Lagos}, {Cerd{\'a}-Dur{\'a}n}, {Davis}, {Gondek-Rosi{\'n}ska}, {Klimenko} et~al.}}]{Szczepanczyk2023}
\bibinfo{author}{\bibfnamefont{M.~J.} \bibnamefont{{Szczepa{\'n}czyk}}}, \bibinfo{author}{\bibfnamefont{Y.}~\bibnamefont{{Zheng}}}, \bibinfo{author}{\bibfnamefont{J.~M.} \bibnamefont{{Antelis}}}, \bibinfo{author}{\bibfnamefont{M.}~\bibnamefont{{Benjamin}}}, \bibinfo{author}{\bibfnamefont{M.-A.} \bibnamefont{{Bizouard}}}, \bibinfo{author}{\bibfnamefont{A.}~\bibnamefont{{Casallas-Lagos}}}, \bibinfo{author}{\bibfnamefont{P.}~\bibnamefont{{Cerd{\'a}-Dur{\'a}n}}}, \bibinfo{author}{\bibfnamefont{D.}~\bibnamefont{{Davis}}}, \bibinfo{author}{\bibfnamefont{D.}~\bibnamefont{{Gondek-Rosi{\'n}ska}}}, \bibinfo{author}{\bibfnamefont{S.}~\bibnamefont{{Klimenko}}}, \bibnamefont{et~al.}, \bibinfo{journal}{arXiv e-prints} \bibinfo{eid}{arXiv:2305.16146} (\bibinfo{year}{2023}), \eprint{2305.16146}.

\bibitem[{\citenamefont{{Abbott} et~al.}(2018)\citenamefont{{Abbott}, {Abbott}, {Kagra Collaboration}, and {VIRGO Collaboration}}}]{Abbott:2020_prospects}
\bibinfo{author}{\bibfnamefont{B.~P.} \bibnamefont{{Abbott}}}, \bibinfo{author}{\bibfnamefont{R.}~\bibnamefont{{Abbott}}}, \bibinfo{author}{\bibfnamefont{L.~S.~C.} \bibnamefont{{Kagra Collaboration}}}, \bibnamefont{and} \bibinfo{author}{\bibnamefont{{VIRGO Collaboration}}}, \bibinfo{journal}{Living Rev. in Relativity} \textbf{\bibinfo{volume}{21}}, \bibinfo{eid}{3} (\bibinfo{year}{2018}), \eprint{1304.0670}.

\bibitem[{\citenamefont{{Szczepa{\'n}czyk} et~al.}(2021)\citenamefont{{Szczepa{\'n}czyk}, {Antelis}, {Benjamin}, {Cavagli{\`a}}, {Gondek-Rosi{\'n}ska}, {Hansen}, {Klimenko}, {Morales}, {Moreno}, {Mukherjee} et~al.}}]{Szczepanczyk2021}
\bibinfo{author}{\bibfnamefont{M.~J.} \bibnamefont{{Szczepa{\'n}czyk}}}, \bibinfo{author}{\bibfnamefont{J.~M.} \bibnamefont{{Antelis}}}, \bibinfo{author}{\bibfnamefont{M.}~\bibnamefont{{Benjamin}}}, \bibinfo{author}{\bibfnamefont{M.}~\bibnamefont{{Cavagli{\`a}}}}, \bibinfo{author}{\bibfnamefont{D.}~\bibnamefont{{Gondek-Rosi{\'n}ska}}}, \bibinfo{author}{\bibfnamefont{T.}~\bibnamefont{{Hansen}}}, \bibinfo{author}{\bibfnamefont{S.}~\bibnamefont{{Klimenko}}}, \bibinfo{author}{\bibfnamefont{M.~D.} \bibnamefont{{Morales}}}, \bibinfo{author}{\bibfnamefont{C.}~\bibnamefont{{Moreno}}}, \bibinfo{author}{\bibfnamefont{S.}~\bibnamefont{{Mukherjee}}}, \bibnamefont{et~al.}, \bibinfo{journal}{Phys. Rev. D} \textbf{\bibinfo{volume}{104}}, \bibinfo{eid}{102002} (\bibinfo{year}{2021}), \eprint{2104.06462}.

\bibitem[{\citenamefont{{Mezzacappa} et~al.}(2023)\citenamefont{{Mezzacappa}, {Marronetti}, {Landfield}, {Lentz}, {Murphy}, {Hix}, {Harris}, {Bruenn}, {Blondin}, {Bronson Messer} et~al.}}]{Mezzacappa2023}
\bibinfo{author}{\bibfnamefont{A.}~\bibnamefont{{Mezzacappa}}}, \bibinfo{author}{\bibfnamefont{P.}~\bibnamefont{{Marronetti}}}, \bibinfo{author}{\bibfnamefont{R.~E.} \bibnamefont{{Landfield}}}, \bibinfo{author}{\bibfnamefont{E.~J.} \bibnamefont{{Lentz}}}, \bibinfo{author}{\bibfnamefont{R.~D.} \bibnamefont{{Murphy}}}, \bibinfo{author}{\bibfnamefont{W.~R.} \bibnamefont{{Hix}}}, \bibinfo{author}{\bibfnamefont{J.~A.} \bibnamefont{{Harris}}}, \bibinfo{author}{\bibfnamefont{S.~W.} \bibnamefont{{Bruenn}}}, \bibinfo{author}{\bibfnamefont{J.~M.} \bibnamefont{{Blondin}}}, \bibinfo{author}{\bibfnamefont{O.~E.} \bibnamefont{{Bronson Messer}}}, \bibnamefont{et~al.}, \bibinfo{journal}{\prd} \textbf{\bibinfo{volume}{107}}, \bibinfo{eid}{043008} (\bibinfo{year}{2023}), \eprint{2208.10643}.

\bibitem[{\citenamefont{{Casallas Lagos} et~al.}(2023)\citenamefont{{Casallas Lagos}, {Antelis}, {Moreno}, {Zanolin}, {Mezzacappa}, and {Szczepa{\'n}czyk}}}]{Lagos2023}
\bibinfo{author}{\bibfnamefont{A.}~\bibnamefont{{Casallas Lagos}}}, \bibinfo{author}{\bibfnamefont{J.~M.} \bibnamefont{{Antelis}}}, \bibinfo{author}{\bibfnamefont{C.}~\bibnamefont{{Moreno}}}, \bibinfo{author}{\bibfnamefont{M.}~\bibnamefont{{Zanolin}}}, \bibinfo{author}{\bibfnamefont{A.}~\bibnamefont{{Mezzacappa}}}, \bibnamefont{and} \bibinfo{author}{\bibfnamefont{M.~J.} \bibnamefont{{Szczepa{\'n}czyk}}}, \bibinfo{journal}{arXiv e-prints} \bibinfo{eid}{arXiv:2304.11498} (\bibinfo{year}{2023}), \eprint{2304.11498}.

\bibitem[{\citenamefont{{Bruel} et~al.}(2023)\citenamefont{{Bruel}, {Bizouard}, {Obergaulinger}, {Maturana-Russel}, {Torres-Forn{\'e}}, {Cerd{\'a}-Dur{\'a}n}, {Christensen}, {Font}, and {Meyer}}}]{Bruel:2023}
\bibinfo{author}{\bibfnamefont{T.}~\bibnamefont{{Bruel}}}, \bibinfo{author}{\bibfnamefont{M.-A.} \bibnamefont{{Bizouard}}}, \bibinfo{author}{\bibfnamefont{M.}~\bibnamefont{{Obergaulinger}}}, \bibinfo{author}{\bibfnamefont{P.}~\bibnamefont{{Maturana-Russel}}}, \bibinfo{author}{\bibfnamefont{A.}~\bibnamefont{{Torres-Forn{\'e}}}}, \bibinfo{author}{\bibfnamefont{P.}~\bibnamefont{{Cerd{\'a}-Dur{\'a}n}}}, \bibinfo{author}{\bibfnamefont{N.}~\bibnamefont{{Christensen}}}, \bibinfo{author}{\bibfnamefont{J.~A.} \bibnamefont{{Font}}}, \bibnamefont{and} \bibinfo{author}{\bibfnamefont{R.}~\bibnamefont{{Meyer}}}, \bibinfo{journal}{\prd} \textbf{\bibinfo{volume}{107}}, \bibinfo{eid}{083029} (\bibinfo{year}{2023}), \eprint{2301.10019}.

\bibitem[{\citenamefont{{Abdikamalov} et~al.}(2014)\citenamefont{{Abdikamalov}, {Gossan}, {DeMaio}, and {Ott}}}]{Abdikamalov:2014}
\bibinfo{author}{\bibfnamefont{E.}~\bibnamefont{{Abdikamalov}}}, \bibinfo{author}{\bibfnamefont{S.}~\bibnamefont{{Gossan}}}, \bibinfo{author}{\bibfnamefont{A.~M.} \bibnamefont{{DeMaio}}}, \bibnamefont{and} \bibinfo{author}{\bibfnamefont{C.~D.} \bibnamefont{{Ott}}}, \bibinfo{journal}{\prd} \textbf{\bibinfo{volume}{90}}, \bibinfo{eid}{044001} (\bibinfo{year}{2014}), \eprint{1311.3678}.

\bibitem[{\citenamefont{{Powell} et~al.}(2016)\citenamefont{{Powell}, {Gossan}, {Logue}, and {Heng}}}]{Powell2016}
\bibinfo{author}{\bibfnamefont{J.}~\bibnamefont{{Powell}}}, \bibinfo{author}{\bibfnamefont{S.~E.} \bibnamefont{{Gossan}}}, \bibinfo{author}{\bibfnamefont{J.}~\bibnamefont{{Logue}}}, \bibnamefont{and} \bibinfo{author}{\bibfnamefont{I.~S.} \bibnamefont{{Heng}}}, \bibinfo{journal}{\prd} \textbf{\bibinfo{volume}{94}}, \bibinfo{eid}{123012} (\bibinfo{year}{2016}), \eprint{1610.05573}.

\bibitem[{\citenamefont{Richers et~al.}(2017)\citenamefont{Richers, Ott, Abdikamalov, O’Connor, and Sullivan}}]{Richers}
\bibinfo{author}{\bibfnamefont{S.}~\bibnamefont{Richers}}, \bibinfo{author}{\bibfnamefont{C.~D.} \bibnamefont{Ott}}, \bibinfo{author}{\bibfnamefont{E.}~\bibnamefont{Abdikamalov}}, \bibinfo{author}{\bibfnamefont{E.}~\bibnamefont{O’Connor}}, \bibnamefont{and} \bibinfo{author}{\bibfnamefont{C.}~\bibnamefont{Sullivan}}, \bibinfo{journal}{Phys. Rev. D} \textbf{\bibinfo{volume}{95}}, \bibinfo{pages}{063019} (\bibinfo{year}{2017}), \urlprefix\url{https://journals.aps.org/prd/abstract/10.1103/PhysRevD.95.063019}.

\bibitem[{\citenamefont{{Morozova} et~al.}(2018)\citenamefont{{Morozova}, {Radice}, {Burrows}, and {Vartanyan}}}]{Morozova:2018}
\bibinfo{author}{\bibfnamefont{V.}~\bibnamefont{{Morozova}}}, \bibinfo{author}{\bibfnamefont{D.}~\bibnamefont{{Radice}}}, \bibinfo{author}{\bibfnamefont{A.}~\bibnamefont{{Burrows}}}, \bibnamefont{and} \bibinfo{author}{\bibfnamefont{D.}~\bibnamefont{{Vartanyan}}}, \bibinfo{journal}{Astrophys. J.} \textbf{\bibinfo{volume}{861}}, \bibinfo{eid}{10} (\bibinfo{year}{2018}), \eprint{1801.01914}.

\bibitem[{\citenamefont{{Torres-Forn{\'e}} et~al.}(2019)\citenamefont{{Torres-Forn{\'e}}, {Cerd{\'a}-Dur{\'a}n}, {Obergaulinger}, {M{\"u}ller}, and {Font}}}]{TF-universal-2019}
\bibinfo{author}{\bibfnamefont{A.}~\bibnamefont{{Torres-Forn{\'e}}}}, \bibinfo{author}{\bibfnamefont{P.}~\bibnamefont{{Cerd{\'a}-Dur{\'a}n}}}, \bibinfo{author}{\bibfnamefont{M.}~\bibnamefont{{Obergaulinger}}}, \bibinfo{author}{\bibfnamefont{B.}~\bibnamefont{{M{\"u}ller}}}, \bibnamefont{and} \bibinfo{author}{\bibfnamefont{J.~A.} \bibnamefont{{Font}}}, \bibinfo{journal}{\prl} \textbf{\bibinfo{volume}{123}}, \bibinfo{eid}{051102} (\bibinfo{year}{2019}), \eprint{1902.10048}.

\bibitem[{\citenamefont{{Bizouard} et~al.}(2021)\citenamefont{{Bizouard}, {Maturana-Russel}, {Torres-Forn{\'e}}, {Obergaulinger}, {Cerd{\'a}-Dur{\'a}n}, {Christensen}, {Font}, and {Meyer}}}]{Bizouard:2021}
\bibinfo{author}{\bibfnamefont{M.-A.} \bibnamefont{{Bizouard}}}, \bibinfo{author}{\bibfnamefont{P.}~\bibnamefont{{Maturana-Russel}}}, \bibinfo{author}{\bibfnamefont{A.}~\bibnamefont{{Torres-Forn{\'e}}}}, \bibinfo{author}{\bibfnamefont{M.}~\bibnamefont{{Obergaulinger}}}, \bibinfo{author}{\bibfnamefont{P.}~\bibnamefont{{Cerd{\'a}-Dur{\'a}n}}}, \bibinfo{author}{\bibfnamefont{N.}~\bibnamefont{{Christensen}}}, \bibinfo{author}{\bibfnamefont{J.~A.} \bibnamefont{{Font}}}, \bibnamefont{and} \bibinfo{author}{\bibfnamefont{R.}~\bibnamefont{{Meyer}}}, \bibinfo{journal}{Phys. Rev. D} \textbf{\bibinfo{volume}{103}}, \bibinfo{pages}{063006} (\bibinfo{year}{2021}), \urlprefix\url{https://ui.adsabs.harvard.edu/abs/2021PhRvD.103f3006B/abstract}.

\bibitem[{\citenamefont{{Afle} and {Brown}}(2021)}]{Afle:2021}
\bibinfo{author}{\bibfnamefont{C.}~\bibnamefont{{Afle}}} \bibnamefont{and} \bibinfo{author}{\bibfnamefont{D.~A.} \bibnamefont{{Brown}}}, \bibinfo{journal}{\prd} \textbf{\bibinfo{volume}{103}}, \bibinfo{eid}{023005} (\bibinfo{year}{2021}), \eprint{2010.00719}.

\bibitem[{\citenamefont{{Pajkos} et~al.}(2021)\citenamefont{{Pajkos}, {Warren}, {Couch}, {O'Connor}, and {Pan}}}]{Pajkos:2021}
\bibinfo{author}{\bibfnamefont{M.~A.} \bibnamefont{{Pajkos}}}, \bibinfo{author}{\bibfnamefont{M.~L.} \bibnamefont{{Warren}}}, \bibinfo{author}{\bibfnamefont{S.~M.} \bibnamefont{{Couch}}}, \bibinfo{author}{\bibfnamefont{E.~P.} \bibnamefont{{O'Connor}}}, \bibnamefont{and} \bibinfo{author}{\bibfnamefont{K.-C.} \bibnamefont{{Pan}}}, \bibinfo{journal}{Astrophys. J.} \textbf{\bibinfo{volume}{914}}, \bibinfo{eid}{80} (\bibinfo{year}{2021}), \eprint{2011.09000}.

\bibitem[{\citenamefont{{Sotani} et~al.}(2021)\citenamefont{{Sotani}, {Takiwaki}, and {Togashi}}}]{Sotani:2021}
\bibinfo{author}{\bibfnamefont{H.}~\bibnamefont{{Sotani}}}, \bibinfo{author}{\bibfnamefont{T.}~\bibnamefont{{Takiwaki}}}, \bibnamefont{and} \bibinfo{author}{\bibfnamefont{H.}~\bibnamefont{{Togashi}}}, \bibinfo{journal}{\prd} \textbf{\bibinfo{volume}{104}}, \bibinfo{eid}{123009} (\bibinfo{year}{2021}), \eprint{2110.03131}.

\bibitem[{\citenamefont{{Andersen} et~al.}(2021)\citenamefont{{Andersen}, {Zha}, {da Silva Schneider}, {Betranhandy}, {Couch}, and {O'Connor}}}]{Andersen:2021}
\bibinfo{author}{\bibfnamefont{O.~E.} \bibnamefont{{Andersen}}}, \bibinfo{author}{\bibfnamefont{S.}~\bibnamefont{{Zha}}}, \bibinfo{author}{\bibfnamefont{A.}~\bibnamefont{{da Silva Schneider}}}, \bibinfo{author}{\bibfnamefont{A.}~\bibnamefont{{Betranhandy}}}, \bibinfo{author}{\bibfnamefont{S.~M.} \bibnamefont{{Couch}}}, \bibnamefont{and} \bibinfo{author}{\bibfnamefont{E.~P.} \bibnamefont{{O'Connor}}}, \bibinfo{journal}{Astrophys. J.} \textbf{\bibinfo{volume}{923}}, \bibinfo{eid}{201} (\bibinfo{year}{2021}), \eprint{2106.09734}.

\bibitem[{\citenamefont{{Saiz-P{\'e}rez} et~al.}(2022)\citenamefont{{Saiz-P{\'e}rez}, {Torres-Forn{\'e}}, and {Font}}}]{Saiz-Perez2022}
\bibinfo{author}{\bibfnamefont{A.}~\bibnamefont{{Saiz-P{\'e}rez}}}, \bibinfo{author}{\bibfnamefont{A.}~\bibnamefont{{Torres-Forn{\'e}}}}, \bibnamefont{and} \bibinfo{author}{\bibfnamefont{J.~A.} \bibnamefont{{Font}}}, \bibinfo{journal}{Mon. Not. R. Astron. Soc.} \textbf{\bibinfo{volume}{512}}, \bibinfo{pages}{3815} (\bibinfo{year}{2022}), \eprint{2110.12941}.

\bibitem[{\citenamefont{{Pastor-Marcos} et~al.}(2023)\citenamefont{{Pastor-Marcos}, {Cerd{\'a}-Dur{\'a}n}, {Walker}, {Torres-Forn{\'e}}, {Abdikamalov}, {Richers}, and {Font}}}]{Pastor-Marcos:2023}
\bibinfo{author}{\bibfnamefont{C.}~\bibnamefont{{Pastor-Marcos}}}, \bibinfo{author}{\bibfnamefont{P.}~\bibnamefont{{Cerd{\'a}-Dur{\'a}n}}}, \bibinfo{author}{\bibfnamefont{D.}~\bibnamefont{{Walker}}}, \bibinfo{author}{\bibfnamefont{A.}~\bibnamefont{{Torres-Forn{\'e}}}}, \bibinfo{author}{\bibfnamefont{E.}~\bibnamefont{{Abdikamalov}}}, \bibinfo{author}{\bibfnamefont{S.}~\bibnamefont{{Richers}}}, \bibnamefont{and} \bibinfo{author}{\bibfnamefont{J.~A.} \bibnamefont{{Font}}}, \bibinfo{journal}{arXiv e-prints} \bibinfo{eid}{arXiv:2308.03456} (\bibinfo{year}{2023}), \eprint{2308.03456}.

\bibitem[{\citenamefont{Powell et~al.}(2023)\citenamefont{Powell, Iess, Llorens-Monteagudo, Obergaulinger, M\"uller, Torres-Forn\'e, Cuoco, and Font}}]{Powell2023}
\bibinfo{author}{\bibfnamefont{J.}~\bibnamefont{Powell}}, \bibinfo{author}{\bibfnamefont{A.}~\bibnamefont{Iess}}, \bibinfo{author}{\bibfnamefont{M.}~\bibnamefont{Llorens-Monteagudo}}, \bibinfo{author}{\bibfnamefont{M.}~\bibnamefont{Obergaulinger}}, \bibinfo{author}{\bibfnamefont{B.}~\bibnamefont{M\"uller}}, \bibinfo{author}{\bibfnamefont{A.}~\bibnamefont{Torres-Forn\'e}}, \bibinfo{author}{\bibfnamefont{E.}~\bibnamefont{Cuoco}}, \bibnamefont{and} \bibinfo{author}{\bibfnamefont{J.~A.} \bibnamefont{Font}} (\bibinfo{year}{2023}), \eprint{2311.18221}.

\bibitem[{\citenamefont{{Wolfe} et~al.}(2023)\citenamefont{{Wolfe}, {Fr{\"o}hlich}, {Miller}, {Torres-Forn{\'e}}, and {Cerd{\'a}-Dur{\'a}n}}}]{Wolfe:2023}
\bibinfo{author}{\bibfnamefont{N.~E.} \bibnamefont{{Wolfe}}}, \bibinfo{author}{\bibfnamefont{C.}~\bibnamefont{{Fr{\"o}hlich}}}, \bibinfo{author}{\bibfnamefont{J.~M.} \bibnamefont{{Miller}}}, \bibinfo{author}{\bibfnamefont{A.}~\bibnamefont{{Torres-Forn{\'e}}}}, \bibnamefont{and} \bibinfo{author}{\bibfnamefont{P.}~\bibnamefont{{Cerd{\'a}-Dur{\'a}n}}}, \bibinfo{journal}{Astrophys. J.} \textbf{\bibinfo{volume}{954}}, \bibinfo{eid}{161} (\bibinfo{year}{2023}), \eprint{2303.16962}.

\bibitem[{\citenamefont{Mitra et~al.}(2023)\citenamefont{Mitra, Shukirgaliyev, Abylkairov, and Abdikamalov}}]{Mitra:2023}
\bibinfo{author}{\bibfnamefont{A.}~\bibnamefont{Mitra}}, \bibinfo{author}{\bibfnamefont{B.}~\bibnamefont{Shukirgaliyev}}, \bibinfo{author}{\bibfnamefont{Y.~S.} \bibnamefont{Abylkairov}}, \bibnamefont{and} \bibinfo{author}{\bibfnamefont{E.}~\bibnamefont{Abdikamalov}}, \bibinfo{journal}{Mon. Not. R. Astron. Soc.} \textbf{\bibinfo{volume}{520}} (\bibinfo{year}{2023}), \urlprefix\url{https://academic.oup.com/mnras/article-abstract/520/2/2473/6989850?redirectedFrom=fulltext&login=false}.

\bibitem[{\citenamefont{{Abbott} et~al.}(2020{\natexlab{b}})\citenamefont{{Abbott}, {Abbott}, {Abbott}, {Abraham}, {LIGO Scientific Collaboration}, and {Virgo Collaboration}}}]{Abbott-guide}
\bibinfo{author}{\bibfnamefont{B.~P.} \bibnamefont{{Abbott}}}, \bibinfo{author}{\bibfnamefont{R.}~\bibnamefont{{Abbott}}}, \bibinfo{author}{\bibfnamefont{T.~D.} \bibnamefont{{Abbott}}}, \bibinfo{author}{\bibfnamefont{S.}~\bibnamefont{{Abraham}}}, \bibinfo{author}{\bibnamefont{{LIGO Scientific Collaboration}}}, \bibnamefont{and} \bibinfo{author}{\bibnamefont{{Virgo Collaboration}}}, \bibinfo{journal}{Classical and Quantum Gravity} \textbf{\bibinfo{volume}{37}}, \bibinfo{eid}{055002} (\bibinfo{year}{2020}{\natexlab{b}}), \eprint{1908.11170}.

\bibitem[{\citenamefont{Drago et~al.}(2021)\citenamefont{Drago, Klimenko, Lazzaro, and et~al}}]{cWB}
\bibinfo{author}{\bibfnamefont{M.}~\bibnamefont{Drago}}, \bibinfo{author}{\bibfnamefont{S.}~\bibnamefont{Klimenko}}, \bibinfo{author}{\bibfnamefont{C.}~\bibnamefont{Lazzaro}}, \bibnamefont{and} \bibinfo{author}{\bibfnamefont{E.~M.} \bibnamefont{et~al}}, \bibinfo{journal}{SoftwareX} \textbf{\bibinfo{volume}{14}}, \bibinfo{pages}{100678} (\bibinfo{year}{2021}), \urlprefix\url{https://www.softxjournal.com/article/S2352-7110(21)00023-6/fulltext}.

\bibitem[{\citenamefont{{R{\"o}ver} et~al.}(2009)\citenamefont{{R{\"o}ver}, {Bizouard}, {Christensen}, {Dimmelmeier}, {Heng}, and {Meyer}}}]{Rover2009}
\bibinfo{author}{\bibfnamefont{C.}~\bibnamefont{{R{\"o}ver}}}, \bibinfo{author}{\bibfnamefont{M.-A.} \bibnamefont{{Bizouard}}}, \bibinfo{author}{\bibfnamefont{N.}~\bibnamefont{{Christensen}}}, \bibinfo{author}{\bibfnamefont{H.}~\bibnamefont{{Dimmelmeier}}}, \bibinfo{author}{\bibfnamefont{I.~S.} \bibnamefont{{Heng}}}, \bibnamefont{and} \bibinfo{author}{\bibfnamefont{R.}~\bibnamefont{{Meyer}}}, \bibinfo{journal}{\prd} \textbf{\bibinfo{volume}{80}}, \bibinfo{eid}{102004} (\bibinfo{year}{2009}), \eprint{0909.1093}.

\bibitem[{\citenamefont{{Logue} et~al.}(2012)\citenamefont{{Logue}, {Ott}, {Heng}, {Kalmus}, and {Scargill}}}]{Logue2012}
\bibinfo{author}{\bibfnamefont{J.}~\bibnamefont{{Logue}}}, \bibinfo{author}{\bibfnamefont{C.~D.} \bibnamefont{{Ott}}}, \bibinfo{author}{\bibfnamefont{I.~S.} \bibnamefont{{Heng}}}, \bibinfo{author}{\bibfnamefont{P.}~\bibnamefont{{Kalmus}}}, \bibnamefont{and} \bibinfo{author}{\bibfnamefont{J.~H.~C.} \bibnamefont{{Scargill}}}, \bibinfo{journal}{\prd} \textbf{\bibinfo{volume}{86}}, \bibinfo{eid}{044023} (\bibinfo{year}{2012}), \eprint{1202.3256}.

\bibitem[{\citenamefont{{Powell} et~al.}(2017)\citenamefont{{Powell}, {Szczepanczyk}, and {Heng}}}]{Powell2017}
\bibinfo{author}{\bibfnamefont{J.}~\bibnamefont{{Powell}}}, \bibinfo{author}{\bibfnamefont{M.}~\bibnamefont{{Szczepanczyk}}}, \bibnamefont{and} \bibinfo{author}{\bibfnamefont{I.~S.} \bibnamefont{{Heng}}}, \bibinfo{journal}{\prd} \textbf{\bibinfo{volume}{96}}, \bibinfo{eid}{123013} (\bibinfo{year}{2017}), \eprint{1709.00955}.

\bibitem[{\citenamefont{{Green} et~al.}(2020)\citenamefont{{Green}, {Simpson}, and {Gair}}}]{Green:2020}
\bibinfo{author}{\bibfnamefont{S.~R.} \bibnamefont{{Green}}}, \bibinfo{author}{\bibfnamefont{C.}~\bibnamefont{{Simpson}}}, \bibnamefont{and} \bibinfo{author}{\bibfnamefont{J.}~\bibnamefont{{Gair}}}, \bibinfo{journal}{\prd} \textbf{\bibinfo{volume}{102}}, \bibinfo{eid}{104057} (\bibinfo{year}{2020}), \eprint{2002.07656}.

\bibitem[{\citenamefont{Dax et~al.}(2021)\citenamefont{Dax, Green, Gair, Macke, Buonanno, and Schölkopf}}]{dax_real-time_2021}
\bibinfo{author}{\bibfnamefont{M.}~\bibnamefont{Dax}}, \bibinfo{author}{\bibfnamefont{S.~R.} \bibnamefont{Green}}, \bibinfo{author}{\bibfnamefont{J.}~\bibnamefont{Gair}}, \bibinfo{author}{\bibfnamefont{J.~H.} \bibnamefont{Macke}}, \bibinfo{author}{\bibfnamefont{A.}~\bibnamefont{Buonanno}}, \bibnamefont{and} \bibinfo{author}{\bibfnamefont{B.}~\bibnamefont{Schölkopf}}, \bibinfo{journal}{arXiv:2106.12594 [astro-ph, physics:gr-qc]}  (\bibinfo{year}{2021}), \bibinfo{note}{arXiv: 2106.12594}, \urlprefix\url{http://arxiv.org/abs/2106.12594}.

\bibitem[{\citenamefont{Williams et~al.}(2021)\citenamefont{Williams, Veitch, and Messenger}}]{williams_nested_2021}
\bibinfo{author}{\bibfnamefont{M.~J.} \bibnamefont{Williams}}, \bibinfo{author}{\bibfnamefont{J.}~\bibnamefont{Veitch}}, \bibnamefont{and} \bibinfo{author}{\bibfnamefont{C.}~\bibnamefont{Messenger}}, \bibinfo{journal}{Phys. Rev. D} \textbf{\bibinfo{volume}{103}}, \bibinfo{pages}{103006} (\bibinfo{year}{2021}), ISSN \bibinfo{issn}{2470-0010, 2470-0029}, \bibinfo{note}{arXiv:2102.11056 [astro-ph, physics:gr-qc]}, \urlprefix\url{http://arxiv.org/abs/2102.11056}.

\bibitem[{\citenamefont{Bayley et~al.}(2022)\citenamefont{Bayley, Messenger, and Woan}}]{bayley_rapid_2022}
\bibinfo{author}{\bibfnamefont{J.}~\bibnamefont{Bayley}}, \bibinfo{author}{\bibfnamefont{C.}~\bibnamefont{Messenger}}, \bibnamefont{and} \bibinfo{author}{\bibfnamefont{G.}~\bibnamefont{Woan}} (\bibinfo{year}{2022}), \bibinfo{note}{arXiv:2209.02031 [astro-ph]}, \urlprefix\url{http://arxiv.org/abs/2209.02031}.

\bibitem[{\citenamefont{{Gabbard} et~al.}(2022)\citenamefont{{Gabbard}, {Messenger}, {Heng}, {Tonolini}, and {Murray-Smith}}}]{Gabbard:2022}
\bibinfo{author}{\bibfnamefont{H.}~\bibnamefont{{Gabbard}}}, \bibinfo{author}{\bibfnamefont{C.}~\bibnamefont{{Messenger}}}, \bibinfo{author}{\bibfnamefont{I.~S.} \bibnamefont{{Heng}}}, \bibinfo{author}{\bibfnamefont{F.}~\bibnamefont{{Tonolini}}}, \bibnamefont{and} \bibinfo{author}{\bibfnamefont{R.}~\bibnamefont{{Murray-Smith}}}, \bibinfo{journal}{Nature Physics} \textbf{\bibinfo{volume}{18}}, \bibinfo{pages}{112} (\bibinfo{year}{2022}), \eprint{1909.06296}.

\bibitem[{\citenamefont{{Dax} et~al.}(2023)\citenamefont{{Dax}, {Green}, {Gair}, {P{\"u}rrer}, {Wildberger}, {Macke}, {Buonanno}, and {Sch{\"o}lkopf}}}]{Dax:2023}
\bibinfo{author}{\bibfnamefont{M.}~\bibnamefont{{Dax}}}, \bibinfo{author}{\bibfnamefont{S.~R.} \bibnamefont{{Green}}}, \bibinfo{author}{\bibfnamefont{J.}~\bibnamefont{{Gair}}}, \bibinfo{author}{\bibfnamefont{M.}~\bibnamefont{{P{\"u}rrer}}}, \bibinfo{author}{\bibfnamefont{J.}~\bibnamefont{{Wildberger}}}, \bibinfo{author}{\bibfnamefont{J.~H.} \bibnamefont{{Macke}}}, \bibinfo{author}{\bibfnamefont{A.}~\bibnamefont{{Buonanno}}}, \bibnamefont{and} \bibinfo{author}{\bibfnamefont{B.}~\bibnamefont{{Sch{\"o}lkopf}}}, \bibinfo{journal}{\prl} \textbf{\bibinfo{volume}{130}}, \bibinfo{eid}{171403} (\bibinfo{year}{2023}), \eprint{2210.05686}.

\bibitem[{\citenamefont{Bhardwaj et~al.}(2023)\citenamefont{Bhardwaj, Alvey, Miller, Nissanke, and Weniger}}]{bhardwaj2023peregrine}
\bibinfo{author}{\bibfnamefont{U.}~\bibnamefont{Bhardwaj}}, \bibinfo{author}{\bibfnamefont{J.}~\bibnamefont{Alvey}}, \bibinfo{author}{\bibfnamefont{B.~K.} \bibnamefont{Miller}}, \bibinfo{author}{\bibfnamefont{S.}~\bibnamefont{Nissanke}}, \bibnamefont{and} \bibinfo{author}{\bibfnamefont{C.}~\bibnamefont{Weniger}}, \emph{\bibinfo{title}{Peregrine: Sequential simulation-based inference for gravitational wave signals}} (\bibinfo{year}{2023}), \eprint{2304.02035}.

\bibitem[{\citenamefont{{Huerta} et~al.}(2019)\citenamefont{{Huerta}, {Allen}, {Andreoni}, {Antelis}, {Bachelet}, {Berriman}, {Bianco}, {Biswas}, {Carrasco Kind}, {Chard} et~al.}}]{Huerta:2019}
\bibinfo{author}{\bibfnamefont{E.~A.} \bibnamefont{{Huerta}}}, \bibinfo{author}{\bibfnamefont{G.}~\bibnamefont{{Allen}}}, \bibinfo{author}{\bibfnamefont{I.}~\bibnamefont{{Andreoni}}}, \bibinfo{author}{\bibfnamefont{J.~M.} \bibnamefont{{Antelis}}}, \bibinfo{author}{\bibfnamefont{E.}~\bibnamefont{{Bachelet}}}, \bibinfo{author}{\bibfnamefont{G.~B.} \bibnamefont{{Berriman}}}, \bibinfo{author}{\bibfnamefont{F.~B.} \bibnamefont{{Bianco}}}, \bibinfo{author}{\bibfnamefont{R.}~\bibnamefont{{Biswas}}}, \bibinfo{author}{\bibfnamefont{M.}~\bibnamefont{{Carrasco Kind}}}, \bibinfo{author}{\bibfnamefont{K.}~\bibnamefont{{Chard}}}, \bibnamefont{et~al.}, \bibinfo{journal}{Nat. Rev. Phys.} \textbf{\bibinfo{volume}{1}}, \bibinfo{pages}{600} (\bibinfo{year}{2019}), \eprint{1911.11779}.

\bibitem[{\citenamefont{{Cuoco} et~al.}(2020)\citenamefont{{Cuoco}, {Powell}, {Cavagli{\`a}}, {Ackley}, {Bejger}, {Chatterjee}, {Coughlin}, {Coughlin}, {Easter}, {Essick} et~al.}}]{Cuoco:2020}
\bibinfo{author}{\bibfnamefont{E.}~\bibnamefont{{Cuoco}}}, \bibinfo{author}{\bibfnamefont{J.}~\bibnamefont{{Powell}}}, \bibinfo{author}{\bibfnamefont{M.}~\bibnamefont{{Cavagli{\`a}}}}, \bibinfo{author}{\bibfnamefont{K.}~\bibnamefont{{Ackley}}}, \bibinfo{author}{\bibfnamefont{M.}~\bibnamefont{{Bejger}}}, \bibinfo{author}{\bibfnamefont{C.}~\bibnamefont{{Chatterjee}}}, \bibinfo{author}{\bibfnamefont{M.}~\bibnamefont{{Coughlin}}}, \bibinfo{author}{\bibfnamefont{S.}~\bibnamefont{{Coughlin}}}, \bibinfo{author}{\bibfnamefont{P.}~\bibnamefont{{Easter}}}, \bibinfo{author}{\bibfnamefont{R.}~\bibnamefont{{Essick}}}, \bibnamefont{et~al.}, \bibinfo{journal}{arXiv e-prints} \bibinfo{eid}{arXiv:2005.03745} (\bibinfo{year}{2020}), \eprint{2005.03745}.

\bibitem[{\citenamefont{{Zhao} et~al.}(2023)\citenamefont{{Zhao}, {Shi}, {Zhou}, {Cao}, and {Ren}}}]{Zhao:2023}
\bibinfo{author}{\bibfnamefont{T.}~\bibnamefont{{Zhao}}}, \bibinfo{author}{\bibfnamefont{R.}~\bibnamefont{{Shi}}}, \bibinfo{author}{\bibfnamefont{Y.}~\bibnamefont{{Zhou}}}, \bibinfo{author}{\bibfnamefont{Z.}~\bibnamefont{{Cao}}}, \bibnamefont{and} \bibinfo{author}{\bibfnamefont{Z.}~\bibnamefont{{Ren}}}, \bibinfo{journal}{arXiv e-prints} \bibinfo{eid}{arXiv:2311.15585} (\bibinfo{year}{2023}), \eprint{2311.15585}.

\bibitem[{\citenamefont{Stergioulas}(2024)}]{Stergioulas:2024}
\bibinfo{author}{\bibfnamefont{N.}~\bibnamefont{Stergioulas}}, \emph{\bibinfo{title}{Machine learning applications in gravitational wave astronomy}} (\bibinfo{year}{2024}), \eprint{2401.07406}.

\bibitem[{\citenamefont{Álvares and {Font et al.}}(2021)}]{Osvaldo:2021}
\bibinfo{author}{\bibfnamefont{J.~D.} \bibnamefont{Álvares}} \bibnamefont{and} \bibinfo{author}{\bibfnamefont{J.~A.} \bibnamefont{{Font et al.}}}, \bibinfo{journal}{Classical and Quantum Gravity} \textbf{\bibinfo{volume}{38}} (\bibinfo{year}{2021}), \urlprefix\url{https://iopscience.iop.org/article/10.1088/1361-6382/ac0455}.

\bibitem[{\citenamefont{Boudart and Fays}(2022)}]{ALBUS}
\bibinfo{author}{\bibfnamefont{V.}~\bibnamefont{Boudart}} \bibnamefont{and} \bibinfo{author}{\bibfnamefont{M.}~\bibnamefont{Fays}}, in \emph{\bibinfo{booktitle}{2022 IEEE International Conference on Big Data (Big Data)}} (\bibinfo{year}{2022}), pp. \bibinfo{pages}{6599--6601}.

\bibitem[{\citenamefont{Sch\"afer et~al.}(2023)\citenamefont{Sch\"afer, Zelenka, Nitz, Wang, Wu, Guo, Cao, Ren, Nousi, Stergioulas et~al.}}]{mockdatachal}
\bibinfo{author}{\bibfnamefont{M.~B.} \bibnamefont{Sch\"afer}}, \bibinfo{author}{\bibfnamefont{O.~c.~v.} \bibnamefont{Zelenka}}, \bibinfo{author}{\bibfnamefont{A.~H.} \bibnamefont{Nitz}}, \bibinfo{author}{\bibfnamefont{H.}~\bibnamefont{Wang}}, \bibinfo{author}{\bibfnamefont{S.}~\bibnamefont{Wu}}, \bibinfo{author}{\bibfnamefont{Z.-K.} \bibnamefont{Guo}}, \bibinfo{author}{\bibfnamefont{Z.}~\bibnamefont{Cao}}, \bibinfo{author}{\bibfnamefont{Z.}~\bibnamefont{Ren}}, \bibinfo{author}{\bibfnamefont{P.}~\bibnamefont{Nousi}}, \bibinfo{author}{\bibfnamefont{N.}~\bibnamefont{Stergioulas}}, \bibnamefont{et~al.}, \bibinfo{journal}{Phys. Rev. D} \textbf{\bibinfo{volume}{107}}, \bibinfo{pages}{023021} (\bibinfo{year}{2023}), \urlprefix\url{https://link.aps.org/doi/10.1103/PhysRevD.107.023021}.

\bibitem[{\citenamefont{Torres-Forné et~al.}(2016)\citenamefont{Torres-Forné, Marquina, Font, and Ibáñez}}]{torres-forne_denoising_2016}
\bibinfo{author}{\bibfnamefont{A.}~\bibnamefont{Torres-Forné}}, \bibinfo{author}{\bibfnamefont{A.}~\bibnamefont{Marquina}}, \bibinfo{author}{\bibfnamefont{J.~A.} \bibnamefont{Font}}, \bibnamefont{and} \bibinfo{author}{\bibfnamefont{J.~M.} \bibnamefont{Ibáñez}}, \bibinfo{journal}{Phys. Rev. D} \textbf{\bibinfo{volume}{94}}, \bibinfo{pages}{124040} (\bibinfo{year}{2016}), \bibinfo{note}{1612.01305}.

\bibitem[{\citenamefont{Torres-Forné et~al.}(2020)\citenamefont{Torres-Forné, Cuoco, Font, and Marquina}}]{torres-forne_application_2020}
\bibinfo{author}{\bibfnamefont{A.}~\bibnamefont{Torres-Forné}}, \bibinfo{author}{\bibfnamefont{E.}~\bibnamefont{Cuoco}}, \bibinfo{author}{\bibfnamefont{J.~A.} \bibnamefont{Font}}, \bibnamefont{and} \bibinfo{author}{\bibfnamefont{A.}~\bibnamefont{Marquina}}, \bibinfo{journal}{Phys. Rev. D} \textbf{\bibinfo{volume}{102}}, \bibinfo{pages}{023011} (\bibinfo{year}{2020}), \bibinfo{note}{publisher: American Physical Society}, \urlprefix\url{https://link.aps.org/doi/10.1103/PhysRevD.102.023011}.

\bibitem[{\citenamefont{{Liao} and {Lin}}(2021)}]{Liao:2021}
\bibinfo{author}{\bibfnamefont{C.-H.} \bibnamefont{{Liao}}} \bibnamefont{and} \bibinfo{author}{\bibfnamefont{F.-L.} \bibnamefont{{Lin}}}, \bibinfo{journal}{\prd} \textbf{\bibinfo{volume}{103}}, \bibinfo{eid}{124051} (\bibinfo{year}{2021}), \eprint{2101.06685}.

\bibitem[{\citenamefont{Schmidt et~al.}(2021)\citenamefont{Schmidt, Breschi, Gamba, Pagano, Rettegno, Riemenschneider, Bernuzzi, Nagar, and Del~Pozzo}}]{stefano_mlgw}
\bibinfo{author}{\bibfnamefont{S.}~\bibnamefont{Schmidt}}, \bibinfo{author}{\bibfnamefont{M.}~\bibnamefont{Breschi}}, \bibinfo{author}{\bibfnamefont{R.}~\bibnamefont{Gamba}}, \bibinfo{author}{\bibfnamefont{G.}~\bibnamefont{Pagano}}, \bibinfo{author}{\bibfnamefont{P.}~\bibnamefont{Rettegno}}, \bibinfo{author}{\bibfnamefont{G.}~\bibnamefont{Riemenschneider}}, \bibinfo{author}{\bibfnamefont{S.}~\bibnamefont{Bernuzzi}}, \bibinfo{author}{\bibfnamefont{A.}~\bibnamefont{Nagar}}, \bibnamefont{and} \bibinfo{author}{\bibfnamefont{W.}~\bibnamefont{Del~Pozzo}}, \bibinfo{journal}{Phys. Rev. D} \textbf{\bibinfo{volume}{103}}, \bibinfo{pages}{043020} (\bibinfo{year}{2021}), \urlprefix\url{https://link.aps.org/doi/10.1103/PhysRevD.103.043020}.

\bibitem[{\citenamefont{Tissino et~al.}(2023)\citenamefont{Tissino, Carullo, Breschi, Gamba, Schmidt, and Bernuzzi}}]{stefano_bns}
\bibinfo{author}{\bibfnamefont{J.}~\bibnamefont{Tissino}}, \bibinfo{author}{\bibfnamefont{G.}~\bibnamefont{Carullo}}, \bibinfo{author}{\bibfnamefont{M.}~\bibnamefont{Breschi}}, \bibinfo{author}{\bibfnamefont{R.}~\bibnamefont{Gamba}}, \bibinfo{author}{\bibfnamefont{S.}~\bibnamefont{Schmidt}}, \bibnamefont{and} \bibinfo{author}{\bibfnamefont{S.}~\bibnamefont{Bernuzzi}}, \bibinfo{journal}{Phys. Rev. D} \textbf{\bibinfo{volume}{107}}, \bibinfo{pages}{084037} (\bibinfo{year}{2023}), \urlprefix\url{https://link.aps.org/doi/10.1103/PhysRevD.107.084037}.

\bibitem[{\citenamefont{Lopez et~al.}(2022)\citenamefont{Lopez, Boudart, Buijsman, Reza, and Caudill}}]{melissa_gans}
\bibinfo{author}{\bibfnamefont{M.}~\bibnamefont{Lopez}}, \bibinfo{author}{\bibfnamefont{V.}~\bibnamefont{Boudart}}, \bibinfo{author}{\bibfnamefont{K.}~\bibnamefont{Buijsman}}, \bibinfo{author}{\bibfnamefont{A.}~\bibnamefont{Reza}}, \bibnamefont{and} \bibinfo{author}{\bibfnamefont{S.}~\bibnamefont{Caudill}}, \bibinfo{journal}{Phys. Rev. D} \textbf{\bibinfo{volume}{106}}, \bibinfo{pages}{023027} (\bibinfo{year}{2022}), \urlprefix\url{https://link.aps.org/doi/10.1103/PhysRevD.106.023027}.

\bibitem[{\citenamefont{{Lopez} et~al.}(2022)\citenamefont{{Lopez}, {Boudart}, {Schmidt}, and {Caudill}}}]{lopez_gengli}
\bibinfo{author}{\bibfnamefont{M.}~\bibnamefont{{Lopez}}}, \bibinfo{author}{\bibfnamefont{V.}~\bibnamefont{{Boudart}}}, \bibinfo{author}{\bibfnamefont{S.}~\bibnamefont{{Schmidt}}}, \bibnamefont{and} \bibinfo{author}{\bibfnamefont{S.}~\bibnamefont{{Caudill}}}, \bibinfo{journal}{arXiv e-prints} \bibinfo{eid}{arXiv:2205.09204} (\bibinfo{year}{2022}), \eprint{2205.09204}.

\bibitem[{\citenamefont{{Astone} et~al.}(2018)\citenamefont{{Astone}, {Cerd{\'a}-Dur{\'a}n}, {Di Palma}, {Drago}, {Muciaccia}, {Palomba}, and {Ricci}}}]{Astone:2018}
\bibinfo{author}{\bibfnamefont{P.}~\bibnamefont{{Astone}}}, \bibinfo{author}{\bibfnamefont{P.}~\bibnamefont{{Cerd{\'a}-Dur{\'a}n}}}, \bibinfo{author}{\bibfnamefont{I.}~\bibnamefont{{Di Palma}}}, \bibinfo{author}{\bibfnamefont{M.}~\bibnamefont{{Drago}}}, \bibinfo{author}{\bibfnamefont{F.}~\bibnamefont{{Muciaccia}}}, \bibinfo{author}{\bibfnamefont{C.}~\bibnamefont{{Palomba}}}, \bibnamefont{and} \bibinfo{author}{\bibfnamefont{F.}~\bibnamefont{{Ricci}}}, \bibinfo{journal}{\prd} \textbf{\bibinfo{volume}{98}}, \bibinfo{eid}{122002} (\bibinfo{year}{2018}), \eprint{1812.05363}.

\bibitem[{\citenamefont{{Chan} et~al.}(2020)\citenamefont{{Chan}, {Heng}, and {Messenger}}}]{Chan2020}
\bibinfo{author}{\bibfnamefont{M.~L.} \bibnamefont{{Chan}}}, \bibinfo{author}{\bibfnamefont{I.~S.} \bibnamefont{{Heng}}}, \bibnamefont{and} \bibinfo{author}{\bibfnamefont{C.}~\bibnamefont{{Messenger}}}, \bibinfo{journal}{\prd} \textbf{\bibinfo{volume}{102}}, \bibinfo{eid}{043022} (\bibinfo{year}{2020}), \eprint{1912.13517}.

\bibitem[{\citenamefont{{L{\'o}pez} et~al.}(2021)\citenamefont{{L{\'o}pez}, {Di Palma}, {Drago}, {Cerd{\'a}-Dur{\'a}n}, and {Ricci}}}]{Melissa:2021}
\bibinfo{author}{\bibfnamefont{M.}~\bibnamefont{{L{\'o}pez}}}, \bibinfo{author}{\bibfnamefont{I.}~\bibnamefont{{Di Palma}}}, \bibinfo{author}{\bibfnamefont{M.}~\bibnamefont{{Drago}}}, \bibinfo{author}{\bibfnamefont{P.}~\bibnamefont{{Cerd{\'a}-Dur{\'a}n}}}, \bibnamefont{and} \bibinfo{author}{\bibfnamefont{F.}~\bibnamefont{{Ricci}}}, \bibinfo{journal}{Phys. Rev. D} \textbf{\bibinfo{volume}{103}}, \bibinfo{eid}{063011} (\bibinfo{year}{2021}).

\bibitem[{\citenamefont{{Iess} et~al.}(2023)\citenamefont{{Iess}, {Cuoco}, {Morawski}, {Nicolaou}, and {Lahav}}}]{Iess:2023}
\bibinfo{author}{\bibfnamefont{A.}~\bibnamefont{{Iess}}}, \bibinfo{author}{\bibfnamefont{E.}~\bibnamefont{{Cuoco}}}, \bibinfo{author}{\bibfnamefont{F.}~\bibnamefont{{Morawski}}}, \bibinfo{author}{\bibfnamefont{C.}~\bibnamefont{{Nicolaou}}}, \bibnamefont{and} \bibinfo{author}{\bibfnamefont{O.}~\bibnamefont{{Lahav}}}, \bibinfo{journal}{A \& A} \textbf{\bibinfo{volume}{669}}, \bibinfo{eid}{A42} (\bibinfo{year}{2023}), \eprint{2301.09387}.

\bibitem[{\citenamefont{{Sasaoka} et~al.}(2023)\citenamefont{{Sasaoka}, {Koyama}, {Dominguez}, {Sakai}, {Somiya}, {Omae}, and {Takahashi}}}]{Sasaoka:2023}
\bibinfo{author}{\bibfnamefont{S.}~\bibnamefont{{Sasaoka}}}, \bibinfo{author}{\bibfnamefont{N.}~\bibnamefont{{Koyama}}}, \bibinfo{author}{\bibfnamefont{D.}~\bibnamefont{{Dominguez}}}, \bibinfo{author}{\bibfnamefont{Y.}~\bibnamefont{{Sakai}}}, \bibinfo{author}{\bibfnamefont{K.}~\bibnamefont{{Somiya}}}, \bibinfo{author}{\bibfnamefont{Y.}~\bibnamefont{{Omae}}}, \bibnamefont{and} \bibinfo{author}{\bibfnamefont{H.}~\bibnamefont{{Takahashi}}}, \bibinfo{journal}{\prd} \textbf{\bibinfo{volume}{108}}, \bibinfo{eid}{123033} (\bibinfo{year}{2023}), \eprint{2310.09551}.

\bibitem[{\citenamefont{He et~al.}(2015)\citenamefont{He, Zhang, Ren, and Sun}}]{resnet}
\bibinfo{author}{\bibfnamefont{K.}~\bibnamefont{He}}, \bibinfo{author}{\bibfnamefont{X.}~\bibnamefont{Zhang}}, \bibinfo{author}{\bibfnamefont{S.}~\bibnamefont{Ren}}, \bibnamefont{and} \bibinfo{author}{\bibfnamefont{J.}~\bibnamefont{Sun}} (\bibinfo{year}{2015}).

\bibitem[{\citenamefont{Zou et~al.}(2019)\citenamefont{Zou, Wang, Li, and Sheng}}]{rescnn}
\bibinfo{author}{\bibfnamefont{X.}~\bibnamefont{Zou}}, \bibinfo{author}{\bibfnamefont{Z.}~\bibnamefont{Wang}}, \bibinfo{author}{\bibfnamefont{Q.}~\bibnamefont{Li}}, \bibnamefont{and} \bibinfo{author}{\bibfnamefont{W.}~\bibnamefont{Sheng}}, \bibinfo{journal}{Neurocomputing} \textbf{\bibinfo{volume}{367}} (\bibinfo{year}{2019}), \urlprefix\url{https://doi.org/10.1016/j.neucom.2019.08.023}.

\bibitem[{\citenamefont{{Zwerger} and {Mueller}}(1997)}]{Zwerger1997}
\bibinfo{author}{\bibfnamefont{T.}~\bibnamefont{{Zwerger}}} \bibnamefont{and} \bibinfo{author}{\bibfnamefont{E.}~\bibnamefont{{Mueller}}}, \bibinfo{journal}{Astronomy and Astrophysics} \textbf{\bibinfo{volume}{320}}, \bibinfo{pages}{209} (\bibinfo{year}{1997}).

\bibitem[{\citenamefont{Dimmelmeier et~al.}(2002)\citenamefont{Dimmelmeier, Font, and Müller}}]{dimmelmeier2}
\bibinfo{author}{\bibfnamefont{H.}~\bibnamefont{Dimmelmeier}}, \bibinfo{author}{\bibfnamefont{J.~A.} \bibnamefont{Font}}, \bibnamefont{and} \bibinfo{author}{\bibfnamefont{E.}~\bibnamefont{Müller}}, \bibinfo{journal}{Astronomy \& Astrophysics} \textbf{\bibinfo{volume}{393}}, \bibinfo{pages}{523 } (\bibinfo{year}{2002}), \urlprefix\url{https://www.aanda.org/articles/aa/abs/2002/38/aa2576/aa2576.html}.

\bibitem[{\citenamefont{Dimmelmeier et~al.}(2008)\citenamefont{Dimmelmeier, Ott, Marek, and Janka}}]{Dimmelmeier2008}
\bibinfo{author}{\bibfnamefont{H.}~\bibnamefont{Dimmelmeier}}, \bibinfo{author}{\bibfnamefont{C.~D.} \bibnamefont{Ott}}, \bibinfo{author}{\bibfnamefont{A.}~\bibnamefont{Marek}}, \bibnamefont{and} \bibinfo{author}{\bibfnamefont{H.~T.} \bibnamefont{Janka}}, \bibinfo{journal}{Phys. Rev. D} \textbf{\bibinfo{volume}{78}}, \bibinfo{pages}{064056} (\bibinfo{year}{2008}), \eprint{0806.4953}.

\bibitem[{\citenamefont{{Ott} et~al.}(2012)\citenamefont{{Ott}, {Abdikamalov}, {O'Connor}, {Reisswig}, {Haas}, {Kalmus}, {Drasco}, {Burrows}, and {Schnetter}}}]{Ott2012}
\bibinfo{author}{\bibfnamefont{C.~D.} \bibnamefont{{Ott}}}, \bibinfo{author}{\bibfnamefont{E.}~\bibnamefont{{Abdikamalov}}}, \bibinfo{author}{\bibfnamefont{E.}~\bibnamefont{{O'Connor}}}, \bibinfo{author}{\bibfnamefont{C.}~\bibnamefont{{Reisswig}}}, \bibinfo{author}{\bibfnamefont{R.}~\bibnamefont{{Haas}}}, \bibinfo{author}{\bibfnamefont{P.}~\bibnamefont{{Kalmus}}}, \bibinfo{author}{\bibfnamefont{S.}~\bibnamefont{{Drasco}}}, \bibinfo{author}{\bibfnamefont{A.}~\bibnamefont{{Burrows}}}, \bibnamefont{and} \bibinfo{author}{\bibfnamefont{E.}~\bibnamefont{{Schnetter}}}, \bibinfo{journal}{\prd} \textbf{\bibinfo{volume}{86}}, \bibinfo{eid}{024026} (\bibinfo{year}{2012}), \eprint{1204.0512}.

\bibitem[{\citenamefont{{Fuller} et~al.}(2015)\citenamefont{{Fuller}, {Klion}, {Abdikamalov}, and {Ott}}}]{Fuller2015}
\bibinfo{author}{\bibfnamefont{J.}~\bibnamefont{{Fuller}}}, \bibinfo{author}{\bibfnamefont{H.}~\bibnamefont{{Klion}}}, \bibinfo{author}{\bibfnamefont{E.}~\bibnamefont{{Abdikamalov}}}, \bibnamefont{and} \bibinfo{author}{\bibfnamefont{C.~D.} \bibnamefont{{Ott}}}, \bibinfo{journal}{Mon. Not. R. Astron. Soc.} \textbf{\bibinfo{volume}{450}}, \bibinfo{pages}{414} (\bibinfo{year}{2015}), \eprint{1501.06951}.

\bibitem[{\citenamefont{{Edwards}}(2021)}]{Edwards2021}
\bibinfo{author}{\bibfnamefont{M.~C.} \bibnamefont{{Edwards}}}, \bibinfo{journal}{\prd} \textbf{\bibinfo{volume}{103}}, \bibinfo{eid}{024025} (\bibinfo{year}{2021}), \eprint{2009.07367}.

\bibitem[{\citenamefont{O'Connor}(2015)}]{gr1d}
\bibinfo{author}{\bibfnamefont{E.}~\bibnamefont{O'Connor}}, \bibinfo{journal}{The Astrophysical Journal Supplement Series} \textbf{\bibinfo{volume}{219}} (\bibinfo{year}{2015}), \urlprefix\url{https://https://iopscience.iop.org/article/10.1088/0067-0049/219/2/24}.

\bibitem[{\citenamefont{Dimmelmeier et~al.}(2005)\citenamefont{Dimmelmeier, Novak, Font, Ibáñez, and Müller}}]{coconut_paper}
\bibinfo{author}{\bibfnamefont{H.}~\bibnamefont{Dimmelmeier}}, \bibinfo{author}{\bibfnamefont{J.}~\bibnamefont{Novak}}, \bibinfo{author}{\bibfnamefont{J.~A.} \bibnamefont{Font}}, \bibinfo{author}{\bibfnamefont{J.~M.} \bibnamefont{Ibáñez}}, \bibnamefont{and} \bibinfo{author}{\bibfnamefont{E.}~\bibnamefont{Müller}}, \bibinfo{journal}{Phys. Rev. D} \textbf{\bibinfo{volume}{71}}, \bibinfo{pages}{064023} (\bibinfo{year}{2005}), \urlprefix\url{https://journals.aps.org/prd/abstract/10.1103/PhysRevD.71.064023}.

\bibitem[{\citenamefont{Liebendörfer}(2005)}]{liebendorfer}
\bibinfo{author}{\bibfnamefont{M.}~\bibnamefont{Liebendörfer}}, \bibinfo{journal}{The Astrophysical Journal} \textbf{\bibinfo{volume}{633}}, \bibinfo{pages}{1042} (\bibinfo{year}{2005}), \urlprefix\url{https://iopscience.iop.org/article/10.1086/466517}.

\bibitem[{gwp()}]{gwpy}
\emph{\bibinfo{title}{Gwpy}}, \bibinfo{howpublished}{\url{https://gwpy.github.io}}, \bibinfo{note}{accessed in 14-04-2023}.

\bibitem[{\citenamefont{He et~al.}(2016)\citenamefont{He, Zhang, Ren, and Sun}}]{deep_paper_7780459}
\bibinfo{author}{\bibfnamefont{K.}~\bibnamefont{He}}, \bibinfo{author}{\bibfnamefont{X.}~\bibnamefont{Zhang}}, \bibinfo{author}{\bibfnamefont{S.}~\bibnamefont{Ren}}, \bibnamefont{and} \bibinfo{author}{\bibfnamefont{J.}~\bibnamefont{Sun}}, in \emph{\bibinfo{booktitle}{2016 IEEE Conference on Computer Vision and Pattern Recognition (CVPR)}} (\bibinfo{year}{2016}), pp. \bibinfo{pages}{770--778}.

\bibitem[{\citenamefont{Deng et~al.}(2009)\citenamefont{Deng, Dong, Socher, Li, Li, and Fei-Fei}}]{ImageNet}
\bibinfo{author}{\bibfnamefont{J.}~\bibnamefont{Deng}}, \bibinfo{author}{\bibfnamefont{W.}~\bibnamefont{Dong}}, \bibinfo{author}{\bibfnamefont{R.}~\bibnamefont{Socher}}, \bibinfo{author}{\bibfnamefont{L.-J.} \bibnamefont{Li}}, \bibinfo{author}{\bibfnamefont{K.}~\bibnamefont{Li}}, \bibnamefont{and} \bibinfo{author}{\bibfnamefont{L.}~\bibnamefont{Fei-Fei}}, in \emph{\bibinfo{booktitle}{2009 IEEE Conference on Computer Vision and Pattern Recognition}} (\bibinfo{year}{2009}), pp. \bibinfo{pages}{248--255}.

\bibitem[{pyt()}]{python}
\emph{\bibinfo{title}{python}}, \bibinfo{howpublished}{\url{https://www.python.org}}, \bibinfo{note}{accessed in 14-04-2023}.

\bibitem[{sci()}]{scipy}
\emph{\bibinfo{title}{Scipy}}, \bibinfo{howpublished}{\url{https://scipy.org}}, \bibinfo{note}{accessed in 14-04-2023}.

\bibitem[{pyc()}]{pycbc}
\emph{\bibinfo{title}{Pycbc}}, \bibinfo{howpublished}{\url{https://pycbc.org}}, \bibinfo{note}{accessed in 14-04-2023}.

\bibitem[{fas()}]{fastai}
\emph{\bibinfo{title}{fast.ai}}, \bibinfo{howpublished}{\url{https://www.fast.ai}}, \bibinfo{note}{accessed in 14-04-2023}.

\bibitem[{tsa()}]{tsai}
\emph{\bibinfo{title}{tsai: State-of-the-art deep learning library for time series and sequences}}, \bibinfo{howpublished}{\url{https://timeseriesai.github.io/tsai/}}, \bibinfo{note}{accessed in 14-04-2023}.

\bibitem[{\citenamefont{Smith and Topin}(2018)}]{smith17}
\bibinfo{author}{\bibfnamefont{L.~N.} \bibnamefont{Smith}} \bibnamefont{and} \bibinfo{author}{\bibfnamefont{N.}~\bibnamefont{Topin}}, \bibinfo{journal}{arXiv e-prints}  (\bibinfo{year}{2018}), \urlprefix\url{http://arxiv.org/abs/1708.07120}.

\bibitem[{\citenamefont{Smith}(2018)}]{smith18}
\bibinfo{author}{\bibfnamefont{L.~N.} \bibnamefont{Smith}}, \bibinfo{journal}{arXiv e-prints}  (\bibinfo{year}{2018}), \urlprefix\url{http://arxiv.org/abs/1803.09820}.

\bibitem[{\citenamefont{Smith}(2017)}]{smith15}
\bibinfo{author}{\bibfnamefont{L.~N.} \bibnamefont{Smith}}, \bibinfo{journal}{CoRR in arXiv e-prints}  (\bibinfo{year}{2017}), \urlprefix\url{http://arxiv.org/abs/1506.01186}.

\bibitem[{\citenamefont{Gal and Ghahramani}(2016)}]{mcdropout}
\bibinfo{author}{\bibfnamefont{Y.}~\bibnamefont{Gal}} \bibnamefont{and} \bibinfo{author}{\bibfnamefont{Z.}~\bibnamefont{Ghahramani}}, \bibinfo{journal}{arXiv e-prints}  (\bibinfo{year}{2016}), \urlprefix\url{http://arxiv.org/abs/1506.02142}.

\bibitem[{skl()}]{sklearn}
\emph{\bibinfo{title}{scikit-learn: Machine learning in python}}, \bibinfo{howpublished}{\url{https://scikit-learn.org/stable/}}, \bibinfo{note}{accessed in 14-04-2023}.

\bibitem[{dif()}]{different_catalog}
\emph{\bibinfo{title}{Exploring supernova gravitational waves with machine learning}}, \bibinfo{howpublished}{\url{https://zenodo.org/record/7090935}}, \bibinfo{note}{accessed in 14-09-2023}.

\bibitem[{\citenamefont{Steiner et~al.}(2013)\citenamefont{Steiner, Hempel, and Fischer}}]{SF}
\bibinfo{author}{\bibfnamefont{A.~W.} \bibnamefont{Steiner}}, \bibinfo{author}{\bibfnamefont{M.}~\bibnamefont{Hempel}}, \bibnamefont{and} \bibinfo{author}{\bibfnamefont{T.}~\bibnamefont{Fischer}}, \bibinfo{journal}{The Astrophysical Journal} \textbf{\bibinfo{volume}{774}} (\bibinfo{year}{2013}), \urlprefix\url{https://iopscience.iop.org/article/10.1088/0004-637X/774/1/17}.

\end{thebibliography}

\newpage


\end{document}